\documentclass[10pt]{amsart}
\usepackage{stmaryrd}
\usepackage{amsmath,amsfonts,amssymb,color,epsfig,graphics}
\usepackage[colorlinks=true,hyperindex=true]{hyperref}
\usepackage[english]{babel}
\usepackage{enumerate, mathrsfs, mathtools}
\usepackage{enumitem}
\usepackage[a4paper, margin=3.65cm]{geometry}

\usepackage{xcolor}
\allowdisplaybreaks
\usepackage{verbatim}
\usepackage{tikz}
\def\build#1_#2^#3{\mathrel{\mathop{\kern 0pt#1}\limits_{#2}^{#3}}}
\usetikzlibrary{arrows}

\newcommand{\R}{{\mathbb{R}}}
\newcommand{\C}{{\mathbb{C}}}
\newcommand{\Z}{{\mathbb{Z}}}
\newcommand{\N}{{\mathbb{N}}}

\newcommand{\Ac}{\mathcal{A}}
\newcommand{\Bc}{\mathcal{B}}
\newcommand{\Cc}{\mathcal{C}}
\newcommand{\Dc}{\mathcal{D}}
\newcommand{\Ec}{\mathcal{E}}

\newcommand{\Gc}{\mathcal{G}}
\newcommand{\Hc}{\mathcal{H}}
\newcommand{\Ic}{\mathcal{I}}
\newcommand{\Jc}{\mathcal{J}}
\newcommand{\Kc}{\mathcal{K}}
\newcommand{\Nc}{\mathcal{N}}
\newcommand{\Pc}{\mathcal{P}}

\newcommand{\Vc}{\mathcal{V}}

\newcommand{\Gr}{\mathscr{G}}

\newcommand{\un}{{\rm \bf {1}}}

\numberwithin{equation}{section}

\def\id{\mathop{\rm id}\nolimits}

\def\rmi{{\rm i}}
\def \lint{[\![}
\def \rint{]\!]}
\newtheorem{theorem}{Theorem}[section]
\newtheorem{proposition}[theorem]{Proposition}
\newtheorem{lemma}[theorem]{Lemma}
\newtheorem{remark}[theorem]{Remark}
\newtheorem{corollary}[theorem]{Corollary}

\begin{document}

\title[Spectral analysis of the discrete Laplacian]{Spectral analysis of The magnetic Laplacian acting
  on discrete funnels}
\author{Nassim Athmouni}
\address{Universit\'e de Gafsa, Campus Universitaire 2112, Tunisie}
\email{\tt athmouninassim@yahoo.fr}
 \author{Marwa Ennaceur}
 \address{ Department of Mathematics, College of Science, University of Ha'il, Hail 81451, Saudi Arabia}
\email{\tt mar.ennaceur@uoh.edu.sa}
\author{Sylvain Gol\'enia}
\address{Univ. Bordeaux,
Bordeaux INP, CNRS, IMB, UMR 5251, F-33400 Talence, France}
\email{\tt sylvain.golenia@math.u-bordeaux.fr}

\begin{abstract}We study perturbations of the discrete magnetic Laplacian associated to discrete analogs of funnels. We perturb the metric in a long-range way. We establish a propagation estimate and a Limiting Absorption Principle away from possible embedded eigenvalues. The approach is based on a positive commutator technique.
\end{abstract}

\maketitle
\tableofcontents
\section{Introduction}
A finite-volume hyperbolic manifold can be elegantly characterized as the union of three distinct components: a compact region, a cusp, and a funnel. This framework is thoroughly discussed in \cite[Theorem 4.5.7]{Th}. The spectral analysis of the Laplacian associated with a graph is closely intertwined with the geometric properties of the graph. Furthermore, graphs are discretized versions of
manifolds \cite{AEG,GT}.

In \cite{AEG}, it was proved that on a discrete analog of cusps and funnels, the perturbed Laplacian ensures a propagation estimate and a Limiting Absorption Principle away from the possible embedded eigenvalues. These results
remind those in the context of the nature of the essential spectrum. Without trying to be exhaustive, using positive commutator techniques, \cite{S,BoSa} treat the case of $\Z^d$, \cite{AF, GG} study the case of binary trees, \cite{AEGJ} treat the triangular model, \cite{AEGJ2,T2} treat the case of the graphene, \cite{MRT,rits} investigate some general graphs, and \cite{PR,KS} focused on a periodic setting. Some other techniques have been used successfully, e.g., \cite{HN,T1} with some geometric approach and \cite{BrKe}.
In the context of some manifolds of finite volume, \cite{AbTr,GoMo} prove
that the essential spectrum of the (continuous) Laplacian becomes empty under the presence of a magnetic field with compact support. Besides, they establish some Weyl asymptotics. Analogously, for some discrete cusps,
 \cite{GT} classify magnetic potentials that lead to the absence of the essential spectrum and compute a kind of Weyl asymptotic for the magnetic discrete Laplacian. Back to \cite{GoMo}, one also obtains a refined analysis of the spectral measure (propagation estimate, limiting absorption principle) for long-range perturbation of the metric when the essential spectrum occurs relying on a positive commutator technique. We refer to \cite{GoMo} for further comments and references therein.
This paper is a magnetic version of \cite{AEG}. 
Since we are interested only in the presence of continuous spectrum, we focus
on the funnel part, and we treat more general graph as in \cite{AEG}.

To start off, we recall some standard definitions of graph theory. A (non-oriented) \emph{graph} is
a triple
$\Gc:=(\mathcal{E},\mathcal{V}, m, \theta)$, where $\mathcal{V}$ is a finite or countable
set (the \emph{vertices}), $\mathcal{E}:\mathcal{V} \times
\mathcal{V}\rightarrow
\mathbb{R}_{+}$ 
is symmetric, and $m:\Vc\to (0,
\infty)$ is a weight. The \emph{magnetic potential} $\theta:\Vc\times
\Vc\to  {\R/2\pi \Z}$ verifies
$\theta_{x,y}:=\theta(x,y)= -\theta_{y,x}$ and $\theta(x,y):=0$ if
$\Ec(x,y)=0$.
We say that $\Gc$ is \emph{simple} if $m=1$ and $\Ec:\Vc\times \Vc\to
\{0,1\}$.

Given  $x,y\in \mathcal{V}$, we say that $(x,y)$ is an \emph{edge} and that $x$ and $y$ are
\emph{neighbors} if $\mathcal{E}(x,y)>0$. Note that in this case,  since $\mathcal{E}$ is symmetric, $(y,x)$ is also an edge and $y$ and $x$ are neighbors.  We denote this
relationship by $x\sim y$ and the set of neighbors of $x$ by
$\Nc_\Gc(x)$. A graph is \emph{connected},
if for all $x,y\in \Vc$, there exists an $x$-$y$-\emph{path}, i.e.,
there is a finite sequence
\[(x_1,\dotsc,x_{N+1})\in \Vc^{N+1} \mbox{ such that }
x_1=x, \, x_{N+1}=y \mbox{ and } x_n\sim x_{n+1},
\]
for all $n\in\{1,\dotsc,N\}$. The minimal possible $N$ is called the
(unweighted) \emph{distance} between  $x$ and $y$. We denote it by $d(x,y)$.

The space of complex-valued functions acting on the set of vertices $\Vc$ is
denoted by $\Cc(\Vc):=\lbrace f:\Vc \to {\C}\rbrace$. Moreover,  $\Cc_c(\Vc)$ is the subspace of $\Cc(\Vc)$  of functions with finite  support. We consider the Hilbert space
\[\ell^2(\Vc, m):=\left\{ f\in \Cc(\Vc), \quad \sum_{x\in\Vc} m(x)|f(x)|^2 <\infty \right\},\]
endowed with the scalar product, $\langle f,g\rangle:= \sum_{x\in \Vc}m(x) \overline{f(x)}g(x)$.

We define the Hermitian form
\[ Q_{\Gc} (f):= \frac{1}{2}\sum_{x,y \in \Vc} \Ec(x,y) \left| f(x) -e^{\rmi
  \theta_{x,y}}f(y)\right|^2,\]
 for all $f\in \Cc_c(\Vc)$.
The associated \emph{magnetic Laplacian}  is the unique non-negative self-adjoint
operator $\Delta_{\Gc}$  satisfying $\langle f, \Delta_{\Gc} f\rangle_{\ell^2(\Vc,m)}= Q_{\Gc} (f)$, for all $f\in
\Cc_c(\Vc)$.  It is the \emph{Friedrichs extension} of $\Delta_{\Gc}|_{\Cc_c(\Vc)}$, e.g., \cite{CTT3, RS},
 where
\begin{align}\label{e:DeltaG}(\Delta_{\Gc} f)(x)&=\frac{1}{m(x)}\sum_{y\in \Vc}\Ec(x,y) \left(f(x)- e^{\rmi \theta_{x,y}}f(y)\right),\end{align}
for all $f\in \Cc_c(\Vc)$.
We set \[\deg_\Gc(x):=\frac{1}{m(x)}\sum_{y\in \Vc}\Ec(x,y),\]
the degree of $x\in\Vc$. The two objects as linked as follows:
\begin{align}\nonumber
0\leq \langle f, \Delta_{\Gc} f\rangle&= \frac{1}{2} \sum_{x\in \Vc} \sum_{y\sim
  x} \Ec(x,y)|f(x)-e^{\rmi \theta_{x,y}}f(y)|^2
\\
\label{e:majo}
&\leq \sum_{x\in \Vc} \sum_{y\sim
  x} \Ec(x,y)(|f(x)|^2+|f(y)|^2) = 2\langle f, d_\Gc(\cdot) f\rangle,
\end{align}
for $f\in\Cc_c(\Vc)$. In particular, $\Delta_{\Gc}$ is bounded (resp.\ compact) when $d_\Gc(\cdot)$ is bounded (resp.\ compact).

We present a simple version of our model: We consider $\Gc_1:=(\Ec_1, \Vc_1, m_1, \theta
_1)$, where
\[\Vc_1:=\N, \, m_1(n):=e^{n},\, \text{ and } \Ec_1(n,n+1):=e^{(2n+1)/2},\]
for all $n\in\N$ and $\Gc_2:=(\Ec_2,\Vc_2,m_2, \theta_2)$ a connected.
We assume that  $\Vc_2$ is finite or that $\lim_{|x|\to \infty}\deg_{\Gc_2}(x)=0$.
In particular, thanks to
\eqref{e:majo}, we have:
\begin{align}\label{e:Delta2comp}
\Delta_{\Gc_2} \in \Kc(\ell^2(\Vc_2, m_2)),
\end{align} where  $\Kc(\ell^2(\Vc_2, m_2))$ denotes the ideal of  compact operators on $\Kc(\ell^2(\Vc_2, m_2))$.
A novelty in this work is that we allow non-finite $\Vc_2$, unlike in \cite{AEG}.
Let $\Gc:=(\Ec, \Vc, m, \theta)$ be the \emph{twisted cartesian product} $\Gc_1\times_{\tau} \Gc_2$ given by
\begin{align*}\left\{\begin{array}{rl}
m(x,y):=& m_1(x)\times m_2(y),
\\
\Ec\left((x,y),(x',y')\right):=
&\Ec_1(x,x')\times \delta_{y, y'} +\delta_{x,x' } \times \Ec_2(y,y'),
\\
\theta((x,y),(x',y')):=&\theta_1(x,x')\times\delta_{y,y'}+\delta_{x,x'}\times\theta_2(y,y'),\end{array}\right.
\end{align*} for all $x,x'\in\Vc_1$ and $y,y'\in \Vc_2$.\\

If we suppose  that $\Delta_{\Gc_2^{}}$ is compact,it follows that $\Delta_\Gc$
essentially self-adjoint on both $\Cc_c(\Vc)$ and $\ell^2(\Vc_1, m_1)\otimes \Cc_c(\Vc_2)$, see \cite{AEG}. Furthermore
\[\Dc(\Delta_\Gc)= \ell^2(\Vc_1, m_1)\otimes \Dc(1/m_2(\cdot))= \Dc(\deg_\Gc(\cdot)).\]
In particular, The (twisted cartesian) Laplacian $\Delta_{\Gc}$  is a bounded operator if and only if
$\liminf_{|x|\to \infty}m_2(x)>0$, see Corollary \ref{c:domain}.
The form domain of $\Delta_\Gc$ will be of central importance in the following analysis. We define:
\begin{align}\label{e:Gr}
\Gr:=\Dc(\Delta_\Gc^{1/2})= \ell^2(\Vc_1, m_1)\otimes \Dc(1/m_2^{1/2}(\cdot))= \Dc(\deg_\Gc^{1/2}(\cdot)).
\end{align} Throughout the sequel, we denote by $\Gr^{*}$ the dual space of $\Gr$.\\
As well, assuming that the set of \emph{adherent points} of $\bigcup_{ x\in \Vc_2}\{\frac{\alpha}{m_2(x)}, \frac{\beta}{m_2(x)}\}$, is locally finite in $\R$, where
\begin{align}\label{e:c1c2}
\alpha := e^{1/2} + e^{-1/2} -2 \quad \mbox{ and } \quad \beta := e^{1/2} + e^{-1/2} +2,
\end{align}
then the operator $\Delta_{\Gc}$ has no singularly continuous spectrum, see Remark \ref{r:transfert} and Corollary \ref{c:scbarre}, and
 \[\sigma(\Delta_{\Gc})=\sigma_{\rm ac}(\Delta_{\Gc})=\sigma_{\rm ess}(\Delta_{\Gc})= \left[\alpha, \beta \right]\cdot \frac{1}{m_2(\cdot)}.\]We use the notation  $\sigma(\cdot), \sigma_{\rm ac}(\cdot)$ and $\sigma_{\rm ess}(\cdot)$ to denote the set of the spectrum, the set of the absolutely continuous spectrum, and the set of the essential spectrum, respectively.
Note that it is easy to construct an example where the spectrum is $\mathbb{R}_+$.
\begin{figure}\label{figure}
\begin{tikzpicture}[scale=0.55]
\def\test{0.75}
\def\xmax{10}
\begin{scope}[rotate=180]
       \foreach \x in {1,2,...,\xmax}
{\fill[color=black]({(\x)/\test}, 0)circle(.3mm);
\fill[color=black]({(\x)/\test+.25/\test/\x}, 4*0.75/\x)circle(.3mm);
\fill[color=black]({(\x)/\test-1/\test/\x}, 4*1/\x)circle(.3mm);
\draw(\x/\test, 0)--(\x/\test+.25/\test/\x, 4*0.75/\x);
\draw(\x/\test, 0)--(\x/\test-1/\test/\x, 4*1/\x);
\draw(\x/\test+.25/\test/\x, 4*0.75/\x)--(\x/\test-1/\test/\x, 4*1/\x);
\draw(\x/\test, 0)--({(\x+1)/\test}, 0);
\draw(\x/\test-1/\test/\x, 4*1/\x)--({(\x+1)/\test-1/\test/(\x+1)}, {4*1/(\x+1)});
\draw(\x/\test+.25/\test/\x, 4*0.75/\x)--({(\x+1)/\test+.25/\test/(\x+1)}, {4*.75/(\x+1)});
}
\fill[color=black]({(\xmax+1)/\test}, 0)circle(0.3mm);
\fill[color=black]({(\xmax+1)/\test+.25/\test/(\xmax+1)}, {4*0.75/(\xmax+1)})circle(.3mm);
\fill[color=black]({(\xmax+1)/\test-1/\test/(\xmax+1)}, {4*1/(\xmax+1)})circle(.3mm);
\draw({(\xmax+1)/\test}, 0)--({(\xmax+1)/\test+.25/\test/(\xmax+1)}, {4*0.75/(\xmax+1)});
\draw({(\xmax+1)/\test}, 0)--({(\xmax+1)/\test-1/\test/(\xmax+1)}, {4*1/(\xmax+1)});
\draw({(\xmax+1)/\test+.25/\test/(\xmax+1)}, {4*0.75/(\xmax+1)})--({(\xmax+1)/\test-1/\test/(\xmax+1)}, {4*1/(\xmax+1)});
\path(-1, 1) node {$\cdots$};
   \end{scope}
\end{tikzpicture}
\caption{Representation of a funnel}
\end{figure}
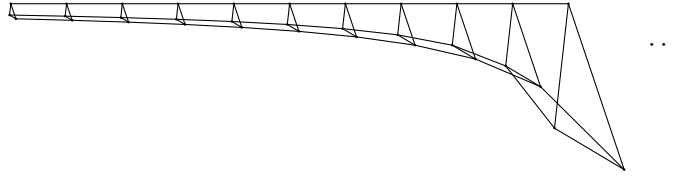

We turn into perturbation theory. First, we perturb the weights, we consider
$\Gc_{\xi,\mu, \gamma}:=(\Ec_\xi,\Vc,m_\mu, \theta_\gamma)$, equipped with a magnetic potential $\theta_\gamma$, where
 \begin{eqnarray*}
m_\mu(x):=&(1+\mu(x))m(x), &\quad \text{ with; } \inf_{x\in \Vc} \mu(x)>-1.
\\
 \Ec_\xi(x,y):=&(1+\xi(x,y))\Ec(x,y), &\quad \text{ with } \inf_{x,y\in \Vc} \xi(x,y)>-1.
\\
\theta_\gamma(x,y):=&(1+\gamma(x,y))\theta(x,y)
 \end{eqnarray*}

 \begin{align*}
  (H_0) \hspace*{1.8 cm }\left\{\begin{array}{rl}
\displaystyle \lim_{|x|\to \infty}\frac{\mu(x)}{1+\deg_\Gc(x)}=0& 
 \\
\displaystyle \lim_{|(x,y)|\to \infty \text{ and } \Ec(x,y)>0}\frac{\xi(x,y)}{1+\deg_\Gc(x)} = 0&
\\
\displaystyle  \lim_{|(x,y)|\to \infty \text{ and }\Ec(x,y)>0} \frac{\gamma(x,y)}{1+\deg_\Gc(x)} = 0&
\\
\displaystyle  \lim_{|x|\to \infty} \frac{V(x)}{1+\deg_\Gc(x)} = 0.&
 \end{array}\right.
\end{align*}

This ensures that $\Delta_{\Gc_{\varepsilon,\mu,\gamma}}$ is a self-adjoint operator with the same form domain as $\Delta_\Gc$.
The domain of the operator is a priori not known. This will lead to further technicalities
in the sequel especially in the treatment of the commutator theory on which we rely.
Moreover, $(H_0)$ guarantees the stability of the essential spectrum, see Proposition \ref{p:com}. Namely,
 \[\sigma_{\rm ess} (\Delta_{\Gc_{\varepsilon,\mu},\gamma})=  \left[\alpha, \beta \right]\cdot \frac{1}{m_2(\cdot)}.\]
In order to obtain the absence of singularly continuous spectrum for $\Delta_{\Gc_{\varepsilon,\mu},\gamma}$,
 we require some additional long-range type decay. From now on, we fix $\epsilon\in(0,1)$ and ask:
 \begin{align*}&(H_1) \sup_{(x_1,x_2)\in\Vc}\langle x_1\rangle^{\epsilon+1} \bigg|\gamma((x_1,x_2),(x_1+1,x_2))-\gamma((x_1,x_2),(x_1-1,x_2) \bigg|(1+\deg_{\Gc}(x_1, x_2))^{-1}\\<\infty,
 \\
 &(H_2) \sup_{(x_1,x_2)\in\Vc} \langle x_1\rangle^{\epsilon+1}|\mu(x_1-1,x_2)-\mu(x_1,x_2)|(1+\deg_{\Gc}(x_1, x_2))^{-1}<\infty,
 \\
 &(H_3) \sup_{(x_1,x_2)\in\Vc} \langle x_1\rangle^{\epsilon+1}|\xi((x_1,x_2),(x_1+1,x_2))-\xi((x_1-1,x_2),(x_1,x_2))|(1+\deg_{\Gc}(x_1, x_2))^{-1}\\<\infty,
  \\
 &(H_4) \sup_{(x_1,x_2)\in\Vc} \langle x_1\rangle^{\epsilon+1}|V(x_1+1,x_2)-V(x_1,x_2)|(1+\deg_{\Gc}(x_1, x_2))^{-1}<\infty,
 \\
 &(H_5) \sup_{(x_1,x_2)\in\Vc}\sup_{y_2\in\Vc_2}\bigg|\gamma((x_1,x_2),(x_1,y_2))\bigg|\deg_{\Gc_2}(y_2)m_2^{\frac{1}{2}}(y_2)<\infty,
 \end{align*}
where $\langle \cdot \rangle := \sqrt{1+|\cdot|^2}$. We stress that we enlarge the class of
perturbation that is given in \cite{AEGJ2}, where the term in $\deg_\Gc$ does not appear. This is due to the fact that we
use of form theoretical approach in this work.
In the sequel, given a function $F(\cdot)$, we denote by $F(Q)$ the operator of multiplication by $F(\cdot)$, i.e., $F(Q)f(x)= F(x)f(x)$.

Our main result is the following:
\begin{theorem}\label{t:LAP} Let $H:= \Delta_{\Gc_{\xi, \mu, \gamma}}+V$ as above. Suppose that $(H_0)$, $(H_1), (H_2)$, $(H_3)$, $(H_4)$, and $(H_5)$ hold true. Set $\kappa(H):=\bigcup_{x\in \Vc_2}\{\frac{\alpha}{m_2(x)}, \frac{\beta}{m_2(x)}\}$ and we assume  $\overline{\kappa(H)}$, is locally finite in $\R$.
 We obtain:
\begin{enumerate}
\item The eigenvalues of $H$ distinct from $\kappa(H)$ are of finite multiplicity and can accumulate
only toward $\kappa(H)$.
\item The singular continuous spectrum of $H$  is empty.
\item Take $s>1/2$ and $[a,b]\subset \R\setminus (\kappa(H)\cup \sigma_{\rm p}(H))$, where $\sigma_{\rm p}$ denotes the point spectrum. The following limit exists and finite:
\[\lim_{\rho\to0^{\pm}}  \sup_{\lambda \in [a,b]}\| (\Lambda|_{\Gr})^{-s}(H-\lambda-\rmi\rho)^{-1} (\Lambda|_{\Gr^{*}})^{-s}\|_{\ell^2(\Vc, m_\mu)}<\infty, \]
where $\Lambda:=\langle Q\rangle\otimes\un_{\Vc_2}$.
Furthermore, in the norm topology of bounded operators, the boundary values of the resolvent:
\[ [a,b] \ni\lambda\mapsto\lim_{\rho\to0^{\pm}}(\Lambda|_{\Gr})^{-s}(H-\lambda-\rmi\rho)^{-1}( \Lambda|_{\Gr^{*}})^{-s} \mbox{ exists and is continuous}.\]
\item There exists $c>0$ such that for all $f\in\ell^2(\Vc, m_\mu)$,  we have:
\[\int_{\R}\|( \Lambda|_{\Gr})^{-s}e^{-\rmi tH}E_{[a,b]}(H)f\|^2_{\ell^2(\Vc, m_\mu)}dt\leq c\|f\|^2_{\ell^2(\Vc, m_\mu)}.\]
\end{enumerate}
\end{theorem}
As in \cite{AEG}, note that we can also glow a finite graph $\Vc$.
The points $(1)-(4)$ are standard consequences of the Mourre's theory. We refer to
Section $2$ for historical references and for an introduction on the subject. This heavy
machinery is a positive commutator technique, which goes in two steps: proving a
Mourre estimate and checking the hypothesis of regularity. The whole Section $4$ is consecrated to it.
Point $(3)$ is called \emph{limiting absorption principle}. Point $(4)$ implies that the spectrum
is purely absolutely continuous above $\R\setminus (\kappa(H)\cup \sigma_{\rm p}(H))$.
Particularly, Riemann-Lebesgue's Theorem ensures that the solution
of the Schr\"{o}dinger equation escapes at infinity. That is, for $f$ belonging to the absolutely continuous
subspace of $H$ and $n\in\Vc$,
\begin{align}\label{l:semi}\lim_{|t|\rightarrow\infty}\big(e^{\rmi t H}f\big)(n)=0.\end{align}
Whereas \eqref{l:semi} can be interpreted as the particle escapes at infinity.
The point $(4)$, corresponds to the fact that $\langle\Lambda\rangle^s$ is locally $H$-smooth over $[a,b]$, see \cite[Section VIII.C]{RS}.

We now describe the structure of the paper. In section 2, we start by construction the graph and, after some short review of commutator properties, we provide an abstract result to establish directly that the operator
is of class $\Cc^k(\Ac)$ (see Propostion \ref{p:Cktens}),  we construct the conjugate operator associated to the unperturbed Laplacian and finish the section by establishing the Mourre estimate. In Section 3, we deal with the perturbation theory and conclude the proof of Theorem \ref{t:LAP}.

\noindent{\bf Notation:} We denote by $\N$ the set of non-negative integers. In particular $0\in \N$. Set $\lint a, b\rint:= [a,b]\cap \Z$.  We denote by $\un_X$ the indicator of the set $X$. We denote by $\Kc(\Hc)$ the ideal of the compact operators of the separable Hilbert space $\Hc$. Given $A,B$ two sets, $A\times B$ is the Cartesian product. Assuming that $A,B\subset \R$, we denote by $A\cdot B:=\{xy, (x,y)\in A\times B \}$.

\section{The free model}
\subsection{Construction of the graph}
\label{s:setting}
We discuss two different product of graphs. To start off,
given $\Gc_1:=(\Ec_1, \Vc_1, m_1, {\theta_1})$ and
$\Gc_2:=(\Ec_2, \Vc_2, m_2, {\theta_2})$, the \emph{Cartesian product of $\Gc_1$ by $\Gc_2$} is defined by
$\Gc^{\diamond}:=(\Ec^{\diamond}, \Vc^{\diamond}, m^{\diamond}, {\theta^\diamond})$, where $\Vc^{\diamond}:= \Vc_1\times \Vc_2$,
\begin{align*}
\left\{\begin{array}{rl}
m^{\diamond}(x,y):=& m_1(x)\times m_2(y),
\\
\Ec^{\diamond}\left((x,y),(x',y')\right):=
&\Ec_1(x,x')\times \delta_{y, y'}  m_2(y)+
m_1(x)\delta_{x,x' } \times \Ec_2(y,y'),
\\
\theta^{\diamond}((x,y),(x',y')):=&\theta_1(x,x')\times\delta_{y,y'}+\delta_{x,x'}\times\theta_2(y,y').
\end{array}\right.
\end{align*}
We denote it by $ \Gc_1\times \Gc_2:= \Gc^{\diamond}$. This definition generalizes the unweighed Cartesian product, e.g., \cite{Ha}. It is used in several places in the literature, e.g., see \cite[Section 2.6]{Ch} and see  \cite{BGKLM} for a generalisation.
\\ The terminology is motivated by the following decomposition:
\begin{align*}
\Delta_{\Gc^{\diamond},\theta^{\diamond}} = \Delta_{\Gc_1,\theta_1} \otimes 1 +
1 \otimes \Delta_{\Gc_2,\theta_2},
\end{align*}
where $\ell^2(\Vc,m)\simeq\ell^2(\Vc_1, m_1)\otimes \ell^2(\Vc_2, m_2)$. Note that
\[e^{\rmi t \Delta_{\Gc^{\diamond},\theta^{\diamond}}} = e^{\rmi t \Delta_{\Gc_1,\theta_1}} \otimes e^{\rmi t \Delta_{\Gc_2,\theta_2}}, \quad \forall t\in \R.\]
 We refer to \cite[Section VIII.10]{RS} for an introduction to
 the tensor product of self-adjoint operators.

We now introduce a \emph{twisted Cartesian product}. We refer to
 \cite[Section 2.2]{GT} for motivations, its link with hyperbolic geometry and generalizations.
Given $\Gc_1:=(\Ec_1, \Vc_1, m_1, {\theta_1})$ and
$\Gc_2:=(\Ec_2, \Vc_2, m_2, {\theta_2})$, we define the \emph{product of $\Gc_1$ by $\Gc_2$} by
$\Gc:=(\Ec, \Vc, m, {\theta})$, where $\Vc:= \Vc_1\times \Vc_2$ and
\begin{align*}
\left\{\begin{array}{rl}
m(x,y):=& m_1(x)\times m_2(y),
\\
\Ec\left((x,y),(x',y')\right):=
&\Ec_1(x,x')\times \delta_{y, y'} +
\delta_{x,x' } \times \Ec_2(y,y'),
\\
\theta((x,y),(x',y')):=&\theta_1(x,x')\times\delta_{y,y'}+\delta_{x,x'}\times\theta_2(y,y'),
\end{array}\right.
\end{align*}
for all $x, x'\in \Vc_1$ and $y, y'\in \Vc_2$. We denote $\Gc$ by
$\Gc_1\times_{\mathfrak{J}} \Gc_2$. If $m=1$,  note that $\Gc_1\times_{\mathfrak{J}} \Gc_2 = \Gc_1\times \Gc_2$.

Under the representation $\ell^2(\Vc, m)\simeq \ell^2(\Vc_1, m_1)\otimes
\ell^2(\Vc_2, m_2)$,
\begin{align}\label{e:deg_cp}
\deg_{\Gc_1\times_{\mathfrak{J}} \Gc_2}(\cdot)= \deg_{\Gc_1}(\cdot)\otimes \frac{1}{m_2(\cdot)}
+ \frac{1}{m_1(\cdot)}\otimes \deg_{\Gc_2}(\cdot)
\end{align}
and
\begin{align}\label{e:rule}
\Delta_{\Gc} = \Delta_{\Gc_1} \otimes \frac{1}{m_2(\cdot)} +
\frac{1}{m_1(\cdot)} \otimes \Delta_{\Gc_2}.
\end{align}
If $m$ is non-trivial, we stress that the Laplacian obtained with our
product is usually not unitarily equivalent to the Laplacian obtained
with the Cartesian product. We refer to \cite[Section VIII.10]{RS} for an introduction to
 the tensor product of self-adjoint operators. We recall the basic results.  For $i\in \{1,2\}$, let $A_i,$ be a self-adjoint operator with domain $\Dc(A_i)$ in the Hilbert spaces $\Hc_i$ and which is essentially self-adjoint on $D_i$. In the Hilbert space $\Hc_1\otimes \Hc_2$, the operators $A_1\otimes A_2$ and $A_1\otimes 1_{\Hc_2} + 1_{\Hc_1} \otimes A_2$ are defined as the closure of $A_1\otimes A_2$ and $A_1\otimes 1_{\Hc_2} + 1_{\Hc_1} \otimes A_2$ acting on $\Dc(A_1)\otimes \Dc(A_2)$.
They are self-adjoint in $\Hc_1\otimes \Hc_2$ and essentially self-adjoint on $D_1\otimes D_2$. They are bounded if and only if $A_1$ and $A_2$ are bounded.  Moreover, $\|A_1\otimes A_2\| = \|A_1\| \cdot \|A_2\|$. Besides, $\sigma(A_1\otimes A_2) = \overline{\sigma(A_1)\cdot \sigma(A_2)}$ and
 $\sigma(A_1\otimes 1_{\Hc_2} + 1_{\Hc_2} \otimes A_2) = \overline{\sigma(A_1)+ \sigma(A_2)}$.

A hyperbolic manifold of finite volume is the union of a compact part, of a cusp, and a funnel, e.g., \cite[Theorem 4.5.7]{Th}. In this article as explain in the introduction, we focus on the discrete funnel part.

Let $n\in\mathbb{N}$ and consider $\Gc_1:=(\Ec_1, \Vc_1, m_1, {\theta_1})$, where
\[\Vc_1:=\N, \quad m_1(n):=\exp(n), \mbox{ and } \Ec_1(n, n+1):= \exp((2n+1)/2),\]
 and $\Gc_2:=(\Ec_2, \Vc_2, m_2, {\theta_2})$ a connected finite graph.
 Set $\Gc^{}:=\Gc_1\times_{\tau} \Gc_2$.
To analyse the perturbations of operator we shall rely on the following gauge transformation, e.g., \cite{Go, CTT, HK}. See also \cite{BG} for some historical references.

To analyze the perturbations of operator we shall rely on the following gauge transformation,
e.g., \cite{Go, CTT, HK}. See also \cite{BG} for some historical references.
\begin{proposition}\label{p:uni}
Let $\Gc:=(\Vc,  \Ec, m, \theta)$ be a weighted graph. Let $m':\Vc\to(0,
\infty)$ be a second  weight. The following map is unitary:
\begin{align}
\nonumber
T_{m\rightarrow m'}f:\ell^2(\Vc, m)&\to \ell^2(\Vc, m')
\\
\label{e:T}
f&\mapsto \left(x \mapsto \sqrt{\frac{m(x)}{m'(x)}} f(x)\right).
\end{align}
In $\ell^2(\Vc, m')$, we have:
\begin{align}\label{e:transform}
\Delta_{\Gc'}=T_{m\rightarrow m'}\left(\Delta_{\Gc}- W(\cdot)\right) T_{m\rightarrow m'}^{-1},
\end{align}
where $\Gc':=(\Vc, \Ec', m', \theta)$ and
\begin{align*}
{\Ec'}(x,y)&:= \Ec (x,y) \sqrt{\frac{m'(x) m'(y)}{m(x) m(y)}}
\\
W(x)&:=\frac{1}{m(x)}\sum_{y\in \Vc} {\Ec}(x,y)  \left(1
 - \sqrt{\frac{m(x)m'(y)}{m(y)m'(x)}}\right).
\end{align*}
\end{proposition}
Note that here, the Laplacians are constructed as Friedrichs extensions.
\subsection{A first step
into commutators}\label{s:mourre}
In the Mourre Theory, one establishes some
spectral properties of a given (unbounded) self-adjoint operator $H$ acting in some complex and separable
Hilbert space $\Hc$ with the help of an external unbounded and self-adjoint operator $\Ac$, by asking a commutator to be positive in a certain sense. We introduce here some properties of
commutators. Let $\|\cdot\|$ denote the norm of bounded operators on $\Hc$. We endow $\Dc(H)$, the domain of $H$, with its graph norm. We denote by $R(z):=(H-z)^{-1}$ the resolvent of $H$ in $z$.
Take an other Hilbert space $\Kc$ such that there is a dense and injective embedding from $\Kc$ to $\Hc$,
by identifying $\Hc$ with its antidual $\Hc^*$, by the Riesz'Lemma,  we have: \begin{equation}\label{inj d}\Kc \hookrightarrow \Hc \simeq \Hc^* \hookrightarrow \Kc^*,\end{equation}
 with  dense and injective embeddings.
We introduce some regularity classes with respect to $\Ac$ and follow \cite[Chapter 6]{ABG}.
that generates a unitary group $W(t) = e^{\rm i t\Ac}$. Given $k\in \N$,  we say that $H\in \Cc^k(\Ac)$ if for all $f\in \Hc$, the map $\R\ni t\mapsto W(t)(H+\rm i)^{-1} W(-t)f\in \Hc$ has the usual $\Cc^k$ regularity.
We say that $H\in \Cc^{k,u}(\Ac)$ if the map $\R\ni t\mapsto W(t)(H+\rmi) W(-t)\in\Bc(\Hc)$ is of class  $\Cc^k(\R,\Bc(\Hc))$, where $\Bc(\Hc)$ is endowed with the norm operator topology. We turn to a criterion in
term of commutator. We begin by introducing  the commutator $[H,\rmi\Ac]$  defined in the form sense on $\Dc(\Ac)\times\Dc(\Ac) $ by $$\langle f,[H,\rmi \Ac]g\rangle:=\rmi\left(\langle Hf,\Ac g\rangle+\langle \Ac f,Hg\rangle\right).$$
By \cite[Lemma 6.2.9]{ABG} we have:
$H\in\Cc^1(A)$ if and only if the form $[H,\rmi A]$ extends to a bounded operator in which case we denote by $[H,\rmi A]_{\circ}$.
In particular, for $H$ being bounded, we obtain $H\Dc(\Ac)\subset \Dc(\Ac)$.
 \begin{theorem}[{\cite[p.251]{ABG}}]  Let $\Ac$ and $H$ be two self-adjoint operators in the Hilbert space $\Hc$.\begin{itemize}
  \item[(a)] \( H \) is of class \( \Cc^1(\Ac) \) if and only if the following two conditions are satisfied:
  \begin{enumerate}
    \item there is a constant \( c < \infty \) such that for all \( f \in \Dc(\Ac) \cap \Dc(H) \):
    \begin{equation}
    \label{eq:6.2.23}
    |\langle \Ac f, Hf \rangle - \langle Hf, \Ac f \rangle| \leq c(\|Hf\|^2 + \|f\|^2),
    \end{equation}
    \item for some \( z \in \mathbb{C} \setminus \sigma(H) \), the set \( \{ f \in \Dc(\Ac) \mid R(z)f \in \Dc(\Ac) \text{ and } R(\bar{z})f \in \Dc(\Ac) \} \) is a core for \( \Ac \).
  \end{enumerate}
  \item[(b)] If \( H \) is of class \( \Cc^1(\Ac) \), then the following is true:
  \begin{enumerate}
    \item[(\(\alpha\))] The space \( R(z)\Dc(\Ac) \) is independent of \( z \in \mathbb{C} \setminus \sigma(H) \) and contained in \( \Dc(\Ac) \), it is a core for \( H \) and a dense subspace of \( \Dc(\Ac) \cap \Dc(H) \) for the intersection topology (i.e., the topology associated to the norm \( \|f\| + \|\Ac f\| + \|Hf\| \));
    \item[(\(\beta\))] The space \( \Dc(\Ac) \cap \Dc(H)\) is a core for \( H \) and the form \( [\Ac, H] \) has a unique extension to a continuous sesquilinear form on \( \Dc(H) \) (equipped with the graph topology); if this extension is denoted by \( [\Ac, H]_{\circ} \), the following identity holds on \( \Hc \) (in the form sense):
    \begin{equation}
    \label{eq:6.2.24}
    [\Ac, R(z)]_{\circ}= -R(z)[\Ac, H]_{\circ}R(z), \qquad z \in \mathbb{C} \setminus \sigma(H).
    \end{equation}
  \end{enumerate}
\end{itemize}
\end{theorem}

\subsection{Construction of the conjugate operator on $\ell^2(\mathbb{N},1)$}

In this section we make a preliminary work for the construction of a conjugate operator for $\Delta_\N$. This is a known result, e.g., \cite{AF}, see also \cite{GG, Mic}.

Given $f\in \ell^{2}(\N,1),$ we set
\[\forall n\in\N^*, \quad Uf(n):=f(n-1) \text{ and } Uf(0):=0.\]
Note that $U^* f(n)=f(n+1), \forall n\in\N$. The operator $U$ is an isometry and is not unitary: we have $U^{*}U=\id$ and $UU^{*}=\un_{[1, \infty[}(\cdot)$.

We define by $Q$ the operator of multiplication by $n$ in $\ell^2(\N,1)$.
Namely, it is the closure of the operator given by  $(Qf)(n)= nf(n)$ for all $n\in\N$ and $f\in \Cc_{c}(\N)$.
It is essentially self-adjoint on $\Cc_{c}(\N)$.
In \cite{GG}, one finds the following elementary relations:
\begin{align}\label{dec}
QU = U(Q+1), \ U^*Q = (Q+1) U^*  \mbox{ and } UQU=U^2(Q+1) \quad \mbox{on } \Dc(Q).\end{align}
The operator $\Delta_{\N}$ is defined by \eqref{e:DeltaG},
 where $\N\simeq(\N,\Ec_{\N},m,0)$, with $\Ec_{\N}(n,n+1)=1$ and $m(n)=1$ for all $n\in\N$.
Explicitly, we have
\[\Delta_\N f(n):=
\left\{\begin{array}{cl}
2f(n)-f(n-1)-f(n+1)& \text{ if } n\geq 1,
\\
f(n)- f(n+1)& \text{ if } n=0,
\end{array}\right.
\quad \forall f\in \ell^2(\N, 1).\]
We can express it with the help of $U$. Namely, we have:
\[\Delta_{\N} = 2 - (U+U^*) - \un_{\{0\}}(\cdot).\]
Using Fourier transform, an easy and standard result is:
\[\sigma_{\rm ess}(\Delta_\N)= [0,4] \quad \mbox{ and } \quad \sigma_{\rm sc}(\Delta_\N)= \emptyset.\]
We construct the conjugate operator in $\ell^2(\N, 1)$. On the space $\Cc_c(\N)$, we define
\begin{align}\nonumber
  \Ac_\N|_{\Cc_c(\N)}&:=\frac{1}{2} \left( SQ +  QS\right), \quad \mbox{ where } S:= \frac{U-U^*}{2\rmi}
  \\ \nonumber
  &=\frac{\rmi}{2}\left(U\left(Q+\frac{1}{2}\right)-U^*\left(Q-\frac{1}{2}\right)\right)
 \\ \label{e:AN}
&=-\frac{\rmi}{2}\left(\frac{1}{2}\left(U^{*}+U\right)+Q\left(U^{*}-U\right)\right).
\end{align}
We denote by $\Ac_\N$ its closure.
\begin{lemma}\label{l:ANessaa}
The operator $\Ac_\N$ is essentially self-adjoint on $\Cc_c(\N)$ and
\[\Dc(\Ac_\N)=\Dc(QS):= \{f\in \ell^2(\N), Sf\in \Dc(Q)\}.\]
\end{lemma}
We refer to \cite{GG} and \cite[Lemma 5.7]{Mic} for the essential self-adjointness and \cite[Lemma 3.1]{GG} for the domain.

We can compute the first commutator. The choice of the $\Ac$ is the first key step to
obtain that the commutator is positive in a certain sense, see \ref{M;E}
\begin{lemma}\label{l:commuA}
The operator $\Delta_\N$ is $\Cc^1(\Ac_\N)$ and we have:
\begin{equation}\label{e:commuA}
[\Delta_{\mathbb{N}},\rmi \Ac_{\mathbb{N}}]_\circ=\frac{1}{2}\Delta_{\mathbb{N}}(4-\Delta_{\mathbb{N}}) + K_1,
\end{equation}
with $K_1$ a finite rank operator belonging to $\Cc^\infty(\Ac)$.
\end{lemma}
This lemma is essentially given in \cite{GG}, see also \cite{AF} for another type of presentation. For the convenience of the reader we reproduce it.
\proof First, since $\delta_{\{0\}}\in \Dc(\Ac^n)$ for all $n\in \N$, $\delta_{\{0\}}$ and $K_1:=[\delta_{\{0\}},\rm i\Ac_{\mathbb{N}}]_\circ$ belong to $\Cc^1(\Ac_\N)$ by  \cite[Proposition 2.1]{GJ1}, see also Lemma \ref{l:class} for a more general result.
Next, we turn to the other part and work in the form sense and by density.
Next, taking $f\in \Cc_c(\N)$ and recalling $\Delta_\N f \in \Cc_c(\N)$, using \eqref{dec} we infer:
 \begin{align*}
 \langle f,[\Delta_{\mathbb{N}},\rmi \Ac_{\mathbb{N}}] f\rangle&:=
 \langle \Delta_\N f, \rmi \Ac_\N f\rangle - \langle -\rmi \Ac_\N f, \Delta_\N f \rangle
\\
&=\rmi\langle f,\Ac_\N(U^{*}+U)-(U^{*}+U)\Ac_{\mathbb{N}} f \rangle +\langle f,[\delta_{\{0\}},\rmi \Ac_{\mathbb{N}}] f \rangle
\\
&= \frac{1}{2}\langle f, \Delta_{\mathbb{N}}(4-\Delta_{\mathbb{N}}) f\rangle +\langle f,[\delta_{\{0\}},\rmi \Ac_{\mathbb{N}}]_\circ f \rangle.
\end{align*}
Since $\Delta_{\mathbb{N}}(4-\Delta_{\mathbb{N}})$ and $[\delta_{\{0\}},\rm i\Ac_{\mathbb{N}}]_\circ$
 are bounded operators and since $ \Cc_c(\N)$ is a core for $\Ac_\N$, there is a constant $c$ such that
 \[|\langle \Delta_\N f, \rmi \Ac_\N f\rangle - \langle -\rmi \Ac_\N f, \Delta_\N f \rangle |\leq c\| f\|^2, \mbox{ for all } f\in \Dc(\Ac).\]
 Hence, it is $\Cc^1(\Ac_\N)$. By density, we also obtain \eqref{e:commuA}.\qed

By induction, since $\Cc^{1}(\Ac)$ is a sub-algebra $\Bc(\Hc)$, we infer:
\begin{corollary} $\Delta_{\N} \in \Cc^\infty(\Ac_\N)$.
\end{corollary}
We mention \cite{Mic} for an anisotropic use on $\Z$ based on the Mourre theory of $\Delta_\N$.

\subsection{Spectral analysis for free model}\label{section3}In this section we construct a conjugate operator for $\Delta_{\Gc^{}}$
 and establish a Mourre estimate.
\subsubsection{First spectral properties}
We introduce the non-magnetic version of $\Gc_1$:
\[\overline{\Gc}_1:=(\overline{\Ec}_1, \overline{\Vc}_1, \overline{m}_1, \overline{\theta}_1):=(\Ec_1, \Vc_1, m_1, 0).\]
Set $\overline{\Gc}^{}:= \overline{\Gc}_1 \times_\tau \Gc_2$.
Recalling the Proposition \ref{p:uni}, we obtain directly:
\begin{align}\label{e:T1DeltaN}
T_{1\rightarrow m_1}^{-1}\Delta_{\overline{\Gc}^{}_1}T_{1\rightarrow m_1}=\Delta_{\N}+ {e^{-1/2}}(e^{1/2}-1)\un_{\{0\}}+
e^{1/2}+e^{-1/2}-2.
\end{align}

As seen above, under the identification
\begin{align}\label{e:otimes}
\ell^{2}(\overline{\Vc}^{},\overline{m}^{})=\ell^{2}(\Vc^{},m^{})&\simeq \ell^{2}(\N,m^{}_1)\otimes  \ell^{2}(\Vc^{}_2, m_2).
\end{align}
We have
\begin{align}\label{e:deltaf}
\Delta_{{\Gc}^{}}= \Delta_{{\Gc}_1^{}} \otimes \frac{1}{m_2^{}(\cdot)} +
\frac{1}{m_1^{}(\cdot)} \otimes  \Delta_{\Gc_2^{}} \mbox{ and } \Delta_{\overline{\Gc}^{}}= \Delta_{\overline{\Gc}_1^{}} \otimes \frac{1}{m_2^{}(\cdot)} +
\frac{1}{m_1^{}(\cdot)} \otimes  \Delta_{\Gc_2^{}}.
\end{align}
As $\N$ is a graph without any triangle,  see also \cite{HiSh, CTT3}, the magnetic Laplacian $\Delta_{\Gc_1}$ is unitarily equivalent to $\Delta_{\overline\Gc_1}$. More precisely, we introduce:
\begin{align}
\nonumber
T_0:\ell^2(\N)&\to \ell^2(\N)
\\
\label{e:T0}
f&\mapsto\left(n \mapsto \left( e^{\rm i \sum_{k=0}^{n-1}\theta_1(k,k+1)} f(n)\right)\right),
\end{align}
with the convention $\sum_{k=0}^{-1}=0$. A direct computation gives:

\begin{align}\label{e:T0Ubis}
T_0\, U T_0^{-1} = e^{\rmi \theta_1(Q-1, Q)} U \text{ and }
T_0\, U^* T_0^{-1} = U^*e^{-\rmi \theta_1(Q-1, Q)}.
\end{align}
With respect to the identification \eqref{e:otimes}, a direct computation gives:
\begin{align}\label{e:uniT0}
\left(T_0 \otimes \id \right) \Delta_{{\Gc}^{}} \left(T_0 \otimes \id \right)^{-1}= \Delta_{{\overline\Gc}^{}}.
\end{align}

Next, note that
\begin{lemma}\label{l:compf} Assume that $\Delta_{\Gc_2^{}}$ is {compact}. Then,
\[\frac{1}{m_1^{}(\cdot)} \otimes  \Delta_{\Gc_2^{}}\in \Kc\left(\ell^2(\Vc^{},m^{})\right).\]
\end{lemma}
\proof Recall that $m_1^{}(n)\to \infty$, as $n\to\infty$. Hence, $\frac{1}{m_1^{}(\cdot)}$ is compact. The result follows by tensorisation. \qed

\begin{remark}
If $\Vc_2$ is finite, $\Delta_{\Gc_2^{}}$ is finite rank hence compact. In case where $\Vc_2$ is infinite. Fix a point a point $x_\circ\in \Vc_2$ and assume $\Gc_2$ connected. Set $|x|:= d(x_\circ, x)$ for $x\in \Vc_2$. Recalling \eqref{e:majo}, by min-max theory, assuming
\[\lim_{|x|\to \infty} \deg_{\Gc_2}(x)= 0,\]
we have $\Delta_{\Gc_2^{}}$ is compact.
\end{remark}
\begin{corollary}\label{c:domain}
Assume that $\Delta_{\Gc_2^{}}$ is {compact}. We have $\Delta_\Gc$ and $\Delta_{\overline \Gc}$
essentially self-adjoint on $\Cc_c(\Vc)$ and $\ell^2(\Vc_1, m_1)\otimes \Cc_c(\Vc_2)$. Moreover
\[\Dc(\Delta_\Gc)= \Dc(\Delta_{\overline \Gc})= \ell^2(\Vc_1, m_1)\otimes \Dc(1/m_2(\cdot))= \Dc(\deg_\Gc(\cdot)).\]
In particular $\Delta_\Gc$ and $\Delta_{\overline \Gc}$ are bounded if and only if
$\inf_{x\in \Vc_2} m_2(x) =0$.
\end{corollary}
\proof For the last equality, note that in \eqref{e:deg_cp}, the map $(x,y)\mapsto (1/m_1(x))\deg_{\Gc_2}(y)$ is bounded. \qed

The form domain of $\Delta_\Gc$ will play a central r\^ole in the sequel. We set
\begin{align}\label{e:Gr}
\Gr:=\Dc(\Delta_\Gc^{1/2})= \Dc(\Delta_{\overline \Gc}^{1/2})= \ell^2(\Vc_1, m_1)\otimes \Dc(1/m_2^{1/2}(\cdot))= \Dc(\deg_\Gc^{1/2}(\cdot)).
\end{align}

We can compute the essential spectrum of $\Delta_{\Gc^{}}$.
 \begin{proposition} Assume that $\Delta_{\Gc_2^{}}$ is compact.
 \[\sigma_{\rm ess}(\Delta_{\Gc^{}})= \left[\alpha, \beta\right]\cdot \frac{1}{m_2(\Vc_2)},\]

 with $\alpha$ and $\beta$ are given in \eqref{e:c1c2}.\
 \end{proposition}

\proof
Recall \eqref{e:T1DeltaN}, Lemma \ref{l:compf} and \eqref{e:uniT0} and use the Weyl theorem. \qed

\begin{remark}
Note that $\Delta_{\Gc^{}}$ is bounded if and only if $\liminf_{|x|\to \infty} m_2(x)=0$. Moreover, it is easy to construct $m_2$ such that $\sigma(\Delta_{\Gc^{}})= \sigma_{\rm ess}(\Delta_{\Gc^{}})=[0, \infty)$.
\end{remark}

\subsubsection{Construction of the conjugate operator on $\ell^2(\Vc,m)$}
In order to analyze the spectral structure of $\Delta_{\Gc^{}}$ and its perturbations, we construct a conjugate operator for $\Delta_{\overline{\Gc}^{}}$. Recalling \eqref{e:AN} and with respect to \eqref{e:otimes}, we set

\begin{align}\label{Afunnel}
\Ac_{\overline{\Gc}^{}}:= 
\Ac_{m_1}\otimes \un_{\Vc_2}:=
T_{1\rightarrow m_1}\Ac_{\N}T^{-1}_{1\rightarrow m_1}\otimes \un_{\Vc_2}.
\end{align}
It is self-adjoint in $\ell^2(\Vc, m)$  and  essentially self-adjoint on $\Cc_c(\Vc^{})$ and on $\Cc_c(\N)\otimes \ell^2(\Vc_2,m_2)$ by Lemma \ref{l:ANessaa}. It acts as follows:

\begin{proposition}\label{p:Am1f} On $\Cc_c(\N)$, we have
 \begin{align*}
 \Ac_{m_1}&=\frac{1}{2}S_{m_1} Q + \text {adjoint}
 \\
&= \frac{\rmi}{2}\left(e^{-1/2}(1/2-Q)U-e^{1/2}(Q+1/2)U^*\right),
 \end{align*}
where $S_{m_1}:= \Im(U_{m_1})$ and $U_{m_1}:= e^{-1/2}U$.
\end{proposition}
\proof For the first line, recalling the first line of \eqref{e:AN}, it is enough to notice that $U_{m_1}= T^{}_{1\rightarrow m_1} UT^{-1}_{1\rightarrow m_1}$. Let us detail the second point and set $f\in\Cc_c(\N)$.
\begin{align*}&\Ac_{m_1}f(n)
=-\frac{\rmi}{2\sqrt{m_1(n)}} \left(\frac{1}{2} \left(U+U^*\right) + Q\left(U^*-U\right)\right)T^{-1}_{1\rightarrow m_1}f(n)
\\
&=\frac{\rmi}{2}\left(\left(n-\frac{1}{2}\right)\sqrt{\frac{m_1(n-1)}{m_1(n)}}f(n-1)-\left(n+\frac{1}{2}\right)
\sqrt{\frac{m_1(n+1)}{m_1(n)}}f(n+1)\right)
\\
&=\frac{\rmi}{2}\left(e^{1/2}\left(n-\frac{1}{2}\right)Uf(n)-e^{-1/2}\left(n+\frac{1}{2}\right)U^*f(n)\right).
\end{align*}
This concludes the proof. \qed\\

We turn to the construction for the conjugate operator for $\Delta_\Gc$. Let $T:=T_{1\to m_1} T_0^{-1}$, where $T_0$ is given
in \eqref{e:T0}. Namely,
\begin{align*}
T:\ell^2(\N, 1)&\to \ell^2(\N, m_1)
\\
f&\mapsto \frac{1}{\sqrt{m_1(n)}}\left( e^{-\rm i\sum_{k=0}^{n-1}\theta_1(k,k+1)} f(n)\right).
\end{align*}
It is a unitary map.
Recalling \eqref{e:AN}, we set the following self-adjoint operator:
\begin{align}\label{e:AG}
\Ac_{\Gc^{}}:=
T\Ac_{\N}T^{-1}\otimes \un_{\Vc_2}.
\end{align}
It is essentially self-adjoint on $\Cc_c(\Vc)$ and on $\Cc_c(\N)\otimes \ell^2(\Vc_2, m_2)$ by Lemma \ref{l:ANessaa}. Thanks to \eqref{e:T0Ubis}, a straightforward computation leads to:

\begin{proposition}\label{p:AG} On $\Cc_c(\Vc)$, we have
 \begin{align*}
\Ac_{\Gc^{}}&=\frac{1}{2}S_{\Gc} Q \otimes \un_{\Vc_2} + \text {adjoint}
 \\
&= \frac{\rm i}{2}\left(e^{-1/2}\big(1/2-Q\big)e^{-\rm i \theta_1(Q-1,Q)}U+e^{1/2}\big(1/2+Q\big)e^{\rm i \theta_1(Q,Q+1)}U^*\right)\otimes \un_{\Vc_2}.
 \end{align*}
with $S_{\Gc}:= \Im(U_{\Gc})$, $U_{\Gc}:= e^{-1/2}e^{-\rm i \theta_1(Q-1,Q)} U$ and $U_{\Gc}^*= e^{1/2}e^{\rm i \theta_1(Q,Q+1)} U^*  $.
\end{proposition}

\begin{remark}\label{r:transfert} Recalling \eqref{e:uniT0}, all the results of Section \ref{section3} are valid for $\Delta_{\Gc^{}}$ and $\Ac_{\Gc^{}}$ in place of $\Delta_{{\overline \Gc}^{}}$ and $\Ac_{{\overline \Gc}^{}}$, especially Proposition \ref{p:Mourrem2} and Corollary \ref{c:scbarre}.
\end{remark}

\subsubsection{Further steps into commutators}\label{S,f}
As the domain is not known a \emph{priori}  in  our case, we rely on a different class of $\Cc^k$ which is
more adapted to the form theory. Let $H$ be a self-adjoint operator
bounded from below and denote by $\Gr$ its form domain, equipped with its natural norm $\|\cdot\|_{\Gr}$. As before, let
$\Ac$ be a self-adjoint operator and set
$W(t):=e^{\rm it\Ac}
$. We ask that $W(t)\Gr\subset \Gr$. For every nonnegative integer $j$, define:
\[
\Gr_j := \left\{ f \in \Gr \;\middle|\; t \mapsto W(t)f \text{ is strongly }\Cc^j \right\}.
\]
In particular, one has
\[
\Gr_1 = \left\{ f \in \Gr \cap \Dc(\Ac) \;\middle|\; \Ac f \in \Gr \right\},
\]
and for $f \in \Gr_1$, we have $\Ac|_{\Gr_1}f = \mathcal{A}f$. Moreover, the operator $W(\cdot)$ leaves each space $\Gr_j$ invariant, and by  duality (see Riesz isomorphism \eqref{inj d}), the same holds for the negative-index spaces $\Gr_{-j}$.

Assume there exists a dense subspace $\mathfrak{K}\subset D(\mathcal{A}) \cap \Gr_{-1}$ invariant under all $W(t)$. Then, by Nelson's Lemma (see \cite[Remark 2.35,]{GGM1}, $\mathfrak{K}$ is a core for $\Ac$ on $\Gr_{-1}$. Therefore, any necessary estimates can be checked on $\mathfrak{K}$.

 The operator  $H$ extends uniquely to a continuous operator:
\[
H: \Gr \to \Gr^*,
\]
and for all $z \in \mathbb{C} \setminus \sigma(H)$, the operator $H - z : \Gr\to \Gr^*$ is a topological isomorphism.

Consequently, the iterated commutator $\operatorname{ad}_{\Ac}^j(H)$  defined recursively as follows:  For each \( j \in \mathbb{N} \),
\[
\mathrm{ad}_\Ac^0(H) := H, \quad \mathrm{ad}_\Ac^{j+1}(H) := [A, \mathrm{ad}_\Ac^j(H)].
\] is well-defined as an element of $\mathcal{B}(\Gr_j, \Gr_{-j}^*)$.

We also have the equivalence:
\[
H \in \Cc^k(\Ac; \Gr, \Gr^*)
\quad \Longleftrightarrow \quad
\operatorname{ad}_{\Ac}^j(H) \in \Bc(\Gr, \Gr^*)
\quad \text{for } j = 0, \dots, k,
\]
here the right hand side means that the operator  $\operatorname{ad}_{\Ac}^j(H)$ extends to a bounded operator of  $\mathcal{B}(\Gr_j, \Gr_{-j}^*)$. Therefore, the notation $H \in \Cc^k(\Ac; \Gr, \Gr^*)$ means that the map $t\mapsto W(t) H W(-t)$ is $k$ times differentiable, with respect of the operator norm topology of $\Bc(\Gr,\Gr^{*})$.

Thanks to \cite[Proposition 5.1.6,]{ABG}, one also has:
\[
H \in \Cc^k(\Ac; \Gr, \Gr^*)
\quad \Longleftrightarrow \quad
R(z) = (H - z)^{-1} \in \Cc^k(\Ac; \Gr^*, \Gr),
\] for some (and hence all)  $z \notin \sigma(H)$. This resolvent condition is strictly stronger than the usual
$R(z) \in \Cc^k(\Ac; \Hc)$, i.e., $H \in \Cc^k(\Ac)$ in the standard sense on $\Hc$. See also \cite[Section 2.5]{GGM1}, for a related discussion.\\ In order to prove regularity, we will be show the following estimate: for each integer $j \geq 0$, there exists a constant $c > 0$ such that
\begin{equation}
\left| \langle f, \operatorname{ad}_{\Ac}^j(H) f \rangle \right|
\leq c \left( \|f\|^2 + \|\Ac f\|^{2}_{\Gr} \right),
\quad \text{for all } f \in \Gr_{j}.
\end{equation}

\subsubsection{Commutator of the full free operator}In our interest, we identify $\Hc:=\ell^2(\Vc, m)$ with $\Hc^*$, the set of anti-linear forms, by the Riesz isomorphism \eqref{inj d}. Recalling \eqref{e:Gr}, this gives:
\begin{align}\label{e:inclusion}
\Dc(\Delta_{\Gc}) \xhookrightarrow{} \Dc(\Delta_{\Gc}^{1/2}):=\Gr \xhookrightarrow{}  \Hc\simeq \Hc^* \xhookrightarrow{}
\Dc(\Delta_{\Gc}^{1/2})^*:=\Gr^* \xhookrightarrow{}  \Dc(\Delta_{\Gc})^*,
\end{align}
where $\xhookrightarrow{}$ is a dense and continuous injection. We see that $\Gr^*$ is the completion of $\Hc$ with respect to the norm
\[\|f\|^2_{\Gr^*}:= \|f \|^2_{\Hc} + \|\deg_{\Gr}^{-1/2}(\cdot) f\|^2_{\Hc}. \]

In our case, we sum up in the following remark:
\begin{remark}\label{r:geneG}
Since $e^{\rmi t \Ac_{\overline\Gc}}\Gr =
\left(e^{\rmi t\Ac_{m_1}}\otimes \un_{\Vc_2}\right) \Gr\subset \Gr$, we denote its generator by $\Ac_{\overline \Gc, \Gr}$
and its domain by $\Dc(\Ac_{\overline\Gc, \Gr})$. Noting that $\Gr= \ell^2(\Vc_1, m_1)\otimes \Dc(1/m_2^{1/2})$, this indicates  $\Dc(\Ac_{\overline\Gc, \Gr})= \Dc(\Ac_{m_1})\otimes \Dc(1/m_2^{1/2})$ and that $\Ac_{\overline{\Gc}, \Gr} = \Ac_{m_1} \otimes \id_{\Dc(1/m_2^{1/2})}$.
Likewise, we also have $(e^{\rmi t \Ac_{\overline \Gc}}|_{\Gr})^*\Gr^* \subset \Gr^*$, by duality.
We denote its generator by $\Ac_{\overline \Gc, \Gr^*}$ and its domain by
$\Dc(\Ac_{\overline \Gc,\Gr^*})$. Recalling that $\Gr^*= \ell^2(\Vc_1, m_1)\otimes \Dc(1/m_2^{1/2})^*$,
we see that $\Dc(\Ac_{\overline \Gc,\Gr^*})= \Dc(\Ac_{m_1})\otimes \Dc(m_2^{1/2})^*$
and that $\Ac_{\overline{\Gc}, \Gr^*} = \Ac_{m_1} \otimes \id_{\Dc(1/m_2^{1/2})^*}$ too.
\end{remark}

\begin{remark}\label{r:comH}
Given $H\in \Bc(\Gr, \Gr^*)$, to establish that $H\in \Cc^1(A_{\overline\Gc}; \Gr, \Gr^*)$,
it is enough to show that  there exists $c>0$ such that
\begin{align}\label{e:comH}
|\langle \Ac_{\overline \Gc} f, H f\rangle - \langle H f, \Ac_{\overline \Gc} f\rangle| \leq c \|f\|_{\Gr}^2.
\end{align}
for all $f\in \Cc_c(\Vc)$. Indeed, first note that $\Cc_c(\Vc)$ is a core for $\Ac_{\overline \Gc}$, the generator of $e^{\rmi t \Ac_{\overline \Gc}}|_{\Gr}$, since the $C_0$-group stabilizes $\Cc_c(\Vc)$, by the Nelson's Lemma, e.g. \cite[Remark 2.35]{GGM1}. Therefore, recalling that
$\Dc(\Ac_{\overline\Gc, \Gr}) = \{f\in \Gr\cap \Dc(\Ac_{\overline \Gc}), \Ac_{\overline \Gc}f \in \Gr\}$, we obtain \eqref{e:comH} for all $f\in \Gr$. In particular $[H, \Ac_{\overline \Gc}]$ extends to a bounded operator in $\Bc(\Gr, \Gr^*)$. We denote it by
$[H, \Ac_{\overline \Gc}]_\circ$.
We also obtain that $H\in \Cc^1(\Ac_{\overline \Gc}; \Gr, \Gr^*)$, e.g.\ \cite[Section 2.5]{GGM1}.
\end{remark}

We turn to the regularity. To lighten the notation, given a graph, we write
\[T_1 \simeq T_2 \text{ if there is  $K:\Cc_c(\Vc) \to \Cc_c(\Vc)$ of finite rank such that } T_1 = T_2 + K. \]
We adapt \cite[Proposition 2.1]{GJ1}  and we obtain:
\begin{lemma}\label{l:class}
Set $T_1$ and $T_2$ in $\Bc(\Gr, \Gr^*)$. Assume that  $T_1\simeq T_2$. Then for all $n\in \N$,
\[T_1\in\Cc^n(\Ac_{\overline{\Gc}}; \Gr, \Gr^*) \Longleftrightarrow T_2\in\Cc^n(\Ac_{\overline{\Gc}};\Gr, \Gr^*).\]
\end{lemma}
\proof Let $K:=T_2-T_1$. By finite induction, it is enough to consider $K=|\delta_x \rangle \langle \delta_y |:= \varphi \mapsto \langle \delta_x, \varphi \rangle \delta_y$, with $x,y\in \Vc$ and $\delta_x(z)= 1$ if $x=z$ and $0$ otherwise.
 We have:
 \[\left[|\delta_x \rangle \langle \delta_y |, \Ac_{\overline \Gc}\right]= |\delta_x \rangle \langle \Ac_{\overline \Gc} \delta_y | - |\Ac_{\overline \Gc} \delta_x \rangle \langle \delta_y |\in\Bc(\Hc)\subset
 \Bc(\Gr, \Gr^*).\]
 Therefore, $K \in \Cc^1(\Ac_{\overline \Gc}; \Gr, \Gr^*)$, e.g.\ \cite[Section 2.5]{GGM1}, and with has the equivalence for $n=1$. For higher $n$, proceed by induction. \qed
\begin{lemma}\label{l:C1Af} Assume that $\Delta_{\Gc_2^{}}$ is compact.
We have $\Delta_{\overline{\Gc}^{}}\in \Cc^1(\Ac_{\overline{\Gc}^{}}; \Gr, \Gr^*)$ and
\begin{align}\label{e:C1Af}
 \left[\Delta_{\overline{\Gc}^{}}, \rmi \Ac_{\overline{\Gc}^{}}\right]_\circ  &
= w(\Delta_{\overline{\Gc}^{}_1}) \otimes \frac{1}{m_2(\cdot)} + K,
\end{align}
where
\[w^{}(x):= \frac{1}{2}\left(x-\alpha\right)\left(\beta-x\right),\]
with $\alpha$ and $\beta$ as in \eqref{e:c1c2} and $K$ is a compact operator.
\end{lemma}
\proof
On  $\Cc_c(\Vc)$,  we obtain
\begin{align} \label{e:C1Afa}
\left[\Delta_{\overline{\Gc}^{}_1}\otimes \frac{1}{m_2(\cdot)}, \rmi \Ac_{\overline{\Gc}^{}}\right]&\simeq\frac{1}{2} (\Delta_{\overline{\Gc}^{}_1}-\alpha) (\beta - \Delta_{\overline{\Gc}^{}_1}) \otimes \frac{1}{m_2(\cdot)}.
\end{align}

Therefore, there is $c>0$ such that for all $f\in \Cc_c(\Vc)$
\begin{align*}
\left|\langle \Ac_{\overline\Gc} f, \Delta_{{\overline\Gc}_1}\otimes \frac{1}{m_2(\cdot)} f\rangle -
\langle \Delta_{\overline{\Gc}^{}_1}\otimes \frac{1}{m_2(\cdot)} f, \Ac_{\overline{\Gc}} f\rangle \right|\leq
c\left(\|f\|^2 + \|\deg_{\overline\Gc}^{1/2} (\cdot) f\|^2 \right).
\end{align*}
Hence by Remark \ref{r:comH}, we infer that $\Delta_{\overline{\Gc}^{}_1}\otimes \frac{1}{m_2(\cdot)}
\in \Cc^1(\Ac_{\overline \Gc}; \Gr, \Gr^*)$.

We turn to the second part of $\Delta_{\overline{\Gc}^{}}$ and compute on $\Cc_c(\Vc)$:
\begin{align}
\nonumber
\left[\frac{1}{m_1(\cdot)} , \rmi \Ac_{m_1}\right] \otimes \Delta_{\Gc_2^{}}&=
T_{1\to m_1}\left[\frac{1}{m_1(\cdot)} , \rmi \Ac_\N\right] T_{1\to m_1}^{-1} \otimes \Delta_{\Gc_2^{}}
\\
\label{e:C1Afb}
&= T_{1\to m_1}\left(\frac{1}{2}(e-1) e^{-Q}\left(Q-\frac{1}{2}\right)U\right) T_{1\to m_1}^{-1} \otimes \Delta_{\Gc_2^{}},
\end{align}
The commutator extends to a compact operator in $\Bc(\Hc)$, since $U$ is a bounded operator, $\lim_{n\to \infty} e^{-n}(n-1/2)=0$ and $\Delta_{\Gc_2^{}}$ is compact. In particular,  by using Remark \ref{r:comH}, we infer that
$\frac{1}{m_1(\cdot)} \otimes \Delta_{\Gc_2^{}} \in \Cc^1(\Ac_{\overline \Gc}; \Gr, \Gr^*)$.  \qed

We define $H\in \Cc^2(\Ac_{\overline \Gc}; \Gr, \Gr^*)$
by asking that $[H,\Ac_{\overline \Gc}]_\circ$ is $\Cc^1(\Ac_{\overline \Gc}; \Gr, \Gr^*)$.
\begin{lemma}\label{l:C2Af}
We have $\Delta_{\overline{\Gc}^{}}\in \Cc^2(\Ac_{\overline \Gc}; \Gr, \Gr^*)$.
\end{lemma}
\proof As above, since $\Cc_c(\Vc)$ is a core for $\Ac_{\overline{\Gc}^{}, \Gr}$ it is enough to prove that the commutator $[[\Delta_{\overline{\Gc}^{}}, \rmi \Ac_{\overline{\Gc}^{}}]_\circ, \rmi \Ac_{\overline{\Gc}^{}}]$, defined initially on $\Cc_c(\Vc)$,  extends to an element of $\Bc(\Gr, \Gr^*)$.

We prove that the right hand side of \eqref{e:C1Af} belongs to $\Cc^1(\Ac_{\overline \Gc}; \Gr, \Gr^*)$.
It composed of $w(\Delta_{\overline{\Gc}^{}_1})\otimes \frac{1}{m_2(\cdot)}$ which is $\Cc^1(\Ac_{\overline \Gc}; \Gr, \Gr^*)$. For the contribution of \eqref{e:C1Afa}, it is composed of a product of bounded operators being in $\Cc^1(\Ac_{m_1})$ on the left of the tensor product and by $1/m_2(\cdot)$ on the right, which is in $\Bc\big(\mathcal{D}\big(1/m_2^{1/2}(\cdot)\big),\mathcal{D}\big(1/m_2^{-1/2}(\cdot)\big)\big)$ and of terms with finite support that are treated by Lemma \ref{l:class}. Concerning  the contribution of \eqref{e:C1Afb}, we repeat the computation above and conclude using $\lim_{n\to \infty} e^{-n}\langle n\rangle^2=0$. We infer it is a compact in $\Bc(\Hc)\subset \Bc(\Gr, \Gr^*)$. Again Remark \ref{r:comH},
ensures the result. \qed

\begin{remark} By induction, we can prove that  $\Delta_{\overline{\Gc}^{}}\in \Cc^\infty(\Ac_{\overline{\Gc}^{}}; \Gr, \Gr^*)$.
\end{remark}
To conclude this section, we give an abstract result that can also be used in order to prove directly that the
operator is of class $\Cc^k(\Ac)$.
\begin{proposition}\label{p:Cktens}
Let $H_1$ and $H_2$ be self-adjoint in $\Hc_1$ and $\Hc_2$ respectively.  Let
$\Ac$ be self-adjoint in $\Hc_1$.
Suppose that $H_1\in \Cc^k(\Ac)$, for some $k\in \N$. Then we have that $H:=H_1\otimes H_2 \in \Cc^k(\Ac\otimes 1)$.
\end{proposition}
\proof Let $z\in\C\setminus\R$, we have
\begin{align*}H-z&=H_1\otimes H_2-z=H_1\otimes H_2-z\id_{\Hc_1}\otimes H_2+z\id_{\Hc_1}\otimes H_2-z
\\
&=(H_1-z)\otimes H_2+z\Big(\id_{\Hc_1}\otimes(H_2-1)\Big).
\end{align*}
Then,
\begin{align*}B:&=(H_1-z)^{-1}\otimes(H_2-z)^{-1}(H-z)
\\
&=\id_{\Hc_1}\otimes H_2(H_2-z)^{-1}
+z(H_1-z)^{-1}\otimes\Big(\id_{\Hc_1}\otimes(H_2-1)\Big)(H_2- z )^{-1},
\end{align*}
is invertible. Note that $B$ is bounded, and $B\in\Cc^1(\Ac\otimes1)$.
Therefore, $B^{-1}\in\Cc^1(\Ac\otimes1)$.
Since $(H-z)^{-1}=B^{-1}(H_1-z)^{-1}\otimes(H_2-z)^{-1}$, and $(H_1-z)^{-1}\otimes(H_2-z)^{-1}\in\Cc^1(\Ac\otimes1)$, then $(H-z)^{-1}\in\Cc^1(\Ac\otimes1)$.
Hence, $(H-z)\in\Cc^1(\Ac\otimes1)$.
\qed
\subsubsection{The Mourre estimate}\label{M;E}
In \cite{Pu}, C.R.\ Putnam used a positive commutator estimate to insure that the spectrum of an operator is purely absolutely continuous. His method was unfortunately not very flexible and did not allow the presence of eigenvalue. In  \cite{Mo81,Mo83}, E.\ Mourre had the idea to localise in energy the positive commutator estimate. Thanks to some hypothesis of regularity, he proved that the embedded eigenvalues can accumulated only at some thresholds, that the singularly continuous spectrum is empty and also established a limiting absorption principle, away from the eigenvalues and from the thresholds. Many papers have shown the power of Mourre's commutator theory for a  wide class of self-adjoint operators, e.g., \cite{BFS,BCHM,CGH,DJ,FH,GGM1,GG,HUS,JMP,S}. We refer to \cite{ABG} for the optimised theory and to \cite{GJ1, GJ,Ge} for recent developments.\\
Given  an interval open interval $\Ic$, we denote by $E_{\Ic}(H)$ the spectral projection of $H$ above $\Ic$. We say that the \emph{Mourre estimate} holds true for a self-adjoint operator $H$ on $\Ic$ if there exist $c>0$ and a compact operator $K$ such that
\begin{align}\label{eq:mourre}
E_\Ic (H)[H, \rmi \Ac]_{\circ}E_\Ic(H)\geq E_\Ic(H)\, (c\, +\, K)\, E_\Ic(H),
\end{align}
when the inequality is understood in the form sense.
We say that we have a \emph{strict Mourre estimate} holds for $H$ on the open interval $\Ic'$ when there exists $c'>0$
such that
\begin{align}\label{eq:mourrestrict}
E_{\Ic'} (H)[H, \rm i \Ac]_{\circ}E_{\Ic'}(H)\geq c' E_{\Ic'}(H).
\end{align}
Assuming $H\in \Cc^1(\Ac)$,  \eqref{eq:mourre}, and $\lambda\in \Ic$ is not an eigenvalue, therefore there exists an open interval $\Ic'$ that contains $\lambda$ and $c'>0$ such that \eqref{eq:mourrestrict}.\\
 This form is essential for obtaining results on the spectral behavior of \( H \), such as \emph{absence of singular spectrum} and the regularity of the continuous spectrum, particularly in the context of \emph{limiting absorption principles (LAP)} see \cite[Theorem 7.6.8]{ABG} or \emph{dispersion inequalities}, which hold true if $H$ is in a slightly better class that $\Cc^{1,u}(\Ac)$.
In this work, our goal is to show a limiting absorption principle.
\begin{proposition}\label{p:Mourrem2}We have $\Delta_{\overline{\Gc}^{}}\in\Cc^2(\Ac_{\overline{\Gc}^{}}; \Gr, \Gr^*)$. Let $\Ic\subset \R$ a compact interval.
Suppose that
\begin{align}\label{e:Mourrem2_1}
\{x\in \Vc_2, m_2(x)\Ic \cap \{\alpha, \beta\}\}= \emptyset
\end{align}
and
\begin{align}\label{e:Mourrem2_2}
\alpha< \inf\{x\in \Vc_2, m_2(x)\Ic \subset (\alpha, \beta)\} \leq
\sup\{x\in \Vc_2, m_2(x)\Ic \subset (\alpha, \beta)\}<\beta.
\end{align}

with $\alpha,\beta$ given in \eqref{e:c1c2}.
Given a compact interval $\Jc$ included in the interior of $\Ic$, there are $c>0$, a compact operator $K$ such that
 \begin{align}
 \label{e:mourref}
 E_\Jc(\Delta_{\overline{\Gc}^{}})[\Delta_{\overline{\Gc}^{}}, \rmi \Ac_{\overline{\Gc}^{}}]_\circ E_\Jc(\Delta_{\overline{\Gc}^{}})&\geq c E_\Jc(\Delta_{\overline{\Gc}^{}}) + K,\end{align} in the form sense. In particular, $\sigma_{\rm sc}(\Delta_{\overline{\Gc}^{}})|_{\Jc}=\emptyset$.
\end{proposition}
\proof
Take $\Ic\subset \R$ a compact interval such that \eqref{e:Mourrem2_1} and \eqref{e:Mourrem2_2} hold true.
First note that
\[\Delta_{\overline{\Gc}^{}_1} \otimes \frac{1}{m_2(\cdot)}=
\bigoplus_{x\in \Vc_2} \frac{1}{m_2(x)} \Delta_{\overline \Gc_1^{}}\otimes \delta_{x}(\cdot) \]
Therefore,  we have
\begin{align*}
E_\Ic\left(\Delta_{\overline{\Gc}^{}_1} \otimes \frac{1}{m_2(\cdot)}\right) &= \bigoplus_{x\in \Vc_2}
E_\Ic\left(\frac{1}{m_2(x)}\Delta_{\overline{\Gc}^{}_1} \right)\otimes\delta_{x}(\cdot)
\\
&= \bigoplus_{x\in \Vc_2}
E_{m_2(x)\Ic} \left(\Delta_{\overline{\Gc}^{}_1} \right)\otimes\delta_{x}(\cdot).
\end{align*}
Let
\begin{align*}
X_{\rm in}(\Ic) &:=\{x\in \Vc_2, m_2(x)\Ic \subset (\alpha, \beta)\}
\\
X_{\rm out}(\Ic) &:=\{x\in \Vc_2, (\alpha, \beta)\cap
m_2(x)\Ic = \emptyset\}.
\end{align*}
Recall that by hypothesis \eqref{e:Mourrem2_1}, we have $\Vc_2= X_{\rm in}(\Ic) \cup X_{\rm out}(\Ic)$.
Note that, as $\sigma \left(\Delta_{\overline{\Gc}^{}_1} \right)= [\alpha, \beta]$,
\[E_{m_2(x)\Ic} \left(\Delta_{\overline{\Gc}^{}_1} \right) = 0 \Longleftrightarrow (\alpha, \beta)\cap
m_2(x)\Ic = \emptyset.\]
We infer:
\begin{align*}
 E_\Ic\left(\Delta_{\overline{\Gc}^{}_1} \otimes \frac{1}{m_2(\cdot)}\right)[\Delta_{\overline{\Gc}^{}}, \rmi \Ac_{\overline{\Gc}^{}}]_\circ E_\Ic\left(
 \Delta_{\overline{\Gc}^{}_1} \otimes \frac{1}{m_2(\cdot)}\right)&
 \\
 &\hspace*{-5cm} \simeq \bigoplus_{x\in \Vc_2} E_{m_2(x)\Ic} \left(\Delta_{\overline{\Gc}^{}_1} \right)w(\Delta_{\overline{\Gc}^{}})\frac{1}{m_2(x)}
 E_{m_2(x)\Ic} \left(\Delta_{\overline{\Gc}^{}_1} \right)\otimes\delta_{x}(\cdot)
 \\
 & \hspace*{-5cm} = \bigoplus_{x\in X_{\rm in}} E_{m_2(x)\Ic} \left(\Delta_{\overline{\Gc}^{}_1} \right)w(\Delta_{\overline{\Gc}^{}})\frac{1}{m_2(x)}
 E_{m_2(x)\Ic} \left(\Delta_{\overline{\Gc}^{}_1} \right)\otimes\delta_{x}(\cdot)
 \\
  & \hspace*{-5cm} \geq \bigoplus_{x\in X_{\rm in}} c
 E_{m_2(x)\Ic} \left(\Delta_{\overline{\Gc}^{}_1} \right)\otimes\delta_{x}(\cdot) =
 \bigoplus_{x\in \Vc_2} c
 E_{m_2(x)\Ic} \left(\Delta_{\overline{\Gc}^{}_1} \right)\otimes\delta_{x}(\cdot)
 \\
 & \hspace*{-5cm} =  c E_{m_2(x)\Ic} \left(\Delta_{\overline{\Gc}^{}_1} \right).
\end{align*}
where $c:= \inf_{x\in X_{\rm in}(\Ic)} \min\left( w (m_2(x) \alpha), w(m_2(x) \beta)\right)\frac{\min \Ic}{\alpha}$ if  $X_{\rm in}(\Ic)\neq \emptyset$ and $c:=1$ if $X_{\rm in}(\Ic)=\emptyset$.
Note that $c>0$ by \eqref{e:Mourrem2_2}.

Next since for all $\varphi\in \Cc^\infty_c(\R)$, by Lemma \ref{l:compf},
\[\varphi \left(\Delta_{\overline{\Gc}^{}_1} \otimes \frac{1}{m_2(\cdot)}\right) - \varphi\left(\Delta_{\overline{\Gc}^{}}\right) \mbox{ is compact,}\]
up to a smaller interval $\Ic\subset \Jc$ and a different compact $K$, we obtain
\begin{align*}
 E_\Ic(\Delta_{\overline{\Gc}^{}})[\Delta_{\overline{\Gc}^{}}, \rmi \Ac_{\overline{\Gc}^{}}]_\circ E_\Ic(\Delta_{\overline{\Gc}^{}})&
 \geq c E_\Ic(\Delta_{\overline{\Gc}^{}}) + K. \end{align*}
The absence of singular continuous spectrum follows from the general theory.  \qed

To lighten the text we did not expand more consequences
 of the Mourre theory in this case and refer to Theorem \ref{t:LAP} for them.

\begin{corollary}\label{c:scbarre}
Assume that the \emph{adherent points} of $\bigcup_{ x\in \Vc_2}\{\frac{\alpha}{m_2(x)}, \frac{\beta}{m_2(x)}\}$:
\begin{align}\label{e:adh}
\overline{\bigcup_{ x\in \Vc_2}\{\frac{\alpha}{m_2(x)}, \frac{\beta}{m_2(x)}\}}\text{ is  locally finite in } \R.
\end{align}
Then $\sigma_{\rm sc}(\Delta_{\overline{\Gc}^{}})= \emptyset$.
\end{corollary}
\proof
 Let $[a,b]\subset \R$. Given $\varepsilon>0$, we set
 $\Pc_\varepsilon:= \bigcup_{ x\in \Vc_2}\{\frac{\alpha}{m_2(x)}, \frac{\beta}{m_2(x)}\} +[-\varepsilon, \varepsilon]$. All $\Ic$ compact intervals in $[a,b]\cap \Pc_{\varepsilon/2}^c$ satisfy the hypothesis of Proposition \ref{p:Mourrem2}. So its conclusion holds true for all compact interval $\Jc$ in $[a,b]\cap \Pc_{\varepsilon}^c$. We infer
 $\sigma_{\rm sc}(\Delta_{\overline{\Gc}^{}})\cap{[a,b]}= \emptyset$, by taking $\varepsilon\to 0$. As $[a,b]$ is arbitrary and \eqref{e:adh}, we get the result.  \qed
\section{The perturbed model}\label{s:perturbed}
In this section, we perturb the metrics of the previous case which will be small to infinity.
\subsection{Perturbation of the metric}\label{Modelp}
Let $\Gc_{\xi, \mu, \gamma}^{}:=(\Vc, \Ec_{\xi}, m_\mu, \theta_\gamma)$ where
\begin{align*}
m_\mu^{}(x)&:=(1+\mu(x))m^{}(x),
\\
 \Ec_\xi^{}(x,y)&:=(1+\xi(x,y))\Ec^{}(x,y),
\\
\theta_\gamma^{}(x,y)&:=\gamma(x,y)+\theta^{}(x,y),
\end{align*}
where
 $\inf \mu>-1$, $\inf \xi >-1$ and satisfying $(H_0)$.
The operator $\Delta_{\Gc_{\xi, \mu, \gamma}^{}}$ is self-adjoint,
as it is a compact perturbation of $\Delta_{\Gc^{}}$, see Proposition \ref{p:com} for a precise statement. In particular,
it has the same essential spectrum as $\Delta_{\Gc^{}}$.

The problem here is that when we apply the perturbation, we do not fall into the same Hilbert space.
To circumvent this obstacle, we rely on Proposition \ref{p:uni} and send unitarily the new Laplacian into $\ell^2(\Vc, m)$. Namely, supposing $(H_0)$, we set
 \begin{align*}
 \widetilde{\Delta}_{\Gc_{\xi, \mu, \gamma}^{}}&:
 = T_{m_{\mu}\rightarrow m}\Delta_{\Gc_{\xi, \mu, \gamma}^{}}T^{-1}_{m_{\mu}\rightarrow m}.
 \end{align*}

A straightforward calculus ensures:

\begin{lemma}For all $f\in \Cc_c(\Vc)$, we have
\begin{align}
\label{e:diff}
&(\widetilde\Delta_{\Gc_{\xi, \mu, \gamma}^{}}-\Delta_{\Gc^{}})f(x)
=
\\
\nonumber&=\frac{1}{m^{}(x)}\sum_{y\in \Vc} \Ec^{}(x,y)\frac{\xi(x,y)}{\sqrt{(1+\mu(y))(1+\mu(x))}}
\Big(f(x)-e^{\rmi \theta_\gamma(x,y)}f(y)\Big)
\\
\nonumber
&\quad-\frac{1}{m^{}(x)}\sum_{y\in \Vc} \Ec^{}(x,y)\left(\frac{\mu(x)+\mu(y)+\mu(x)\mu(y)}{\sqrt{(1+\mu(y))(1+\mu(x))}(1+\sqrt{(1+\mu(y))(1+\mu(x))})}\right)f(x)
\\
&
\nonumber\quad+\frac{1}{m^{}(x)}\sum_{y\in \Vc} \Ec^{}(x,y)\left(e^{\rmi \theta(x,y)}-\frac{e^{\rmi \theta_\gamma(x,y)}}{\sqrt{(1+\mu(y))(1+\mu(x))}}\right)f(y)+Wf(x),
\end{align}
where
\[W(x):=\frac{1}{m_\mu(x)}\sum_{y\in \Vc} {\Ec_\xi}(x,y)  \left(1
 - \sqrt{\frac{m_\mu(x)m(y)}{m_\mu(y)m(x)}}\right).\]
\end{lemma}
We first deal with the question of the essential spectrum. The proof is close to \cite[Proposition 5.2]{Go}.
\begin{proposition}\label{p:com}
We assume that $(H_0)$ holds true
then
\[\Dc(\Delta_{\Gc}^{1/2}) =
\Dc(|\widetilde{\Delta}_{\Gc_{\xi, \mu,\gamma}}+V(\cdot)|^{1/2})= \Dc(\deg_\Gc^{1/2}(\cdot))=
\ell^2(\Vc_1, m_1)\otimes \Dc(1/m_2^{1/2}(\cdot))\]
and
\[(\widetilde{\Delta}_{\Gc_{\xi, \mu,\gamma}^{}}+\rmi)^{-1}-(\Delta_{\Gc^{}}+\rmi)^{-1}\in \Kc(\ell^2(\Vc),m^{}),\] where
 $\widetilde\Delta_{\Gc_{\xi, \mu,\gamma}^{}}:=
T_{m_{\mu}\rightarrow m}\Delta_{\Gc_{\xi, \mu,\gamma}^{}}T^{-1}_{m_{\mu}\rightarrow m}$.
In particular,
\[\sigma_{\rm ess}(\Delta_{\Gc_{\xi, \mu,\gamma}^{}})=\sigma_{\rm ess}(\Delta_{\Gc^{}}).\]
\end{proposition}

\proof Set $H:= \widetilde{\Delta}_{\Gc_{\xi, \mu,\gamma}^{}}+V(\cdot)$. Using \eqref{e:diff} and mimicking \eqref{e:majo}, for all $f\in\Cc_c(\Vc)$, we have
\begin{align*}
&\big|\langle f, (H-\Delta_{\Gc^{}}) f\rangle_{\ell^2(\Vc,m)}\big|
\\
&\leq\sum_{x\in \Vc}\sum_{y\in \Vc} \Ec(x,y)\frac{|\xi(x,y)|}{\sqrt{(1+\mu(y))(1+\mu(x))}}|f(x)- e^{\rmi \theta_\gamma(x,y)}f(y)|^2
\\
&\quad+\langle f,\deg_2(\cdot)+|W(\cdot)|f\rangle_{\ell^2(\Vc,m)}
\\
&\quad+\sum_{x\in \Vc}\sum_{y\in \Vc} \Ec(x,y)\left|e^{\rmi \theta(x,y)}-\frac{e^{\rmi \theta_\gamma(x,y)}}{\sqrt{(1+\mu(y))(1+\mu(x))}}\right| \cdot\left(|f(x)-f(y)|^2+|f(x)|^2\right)
\\
&\leq\langle f,\left(2\deg_1(\cdot)+\deg_2(\cdot)+3\deg_3(\cdot)+|W|(\cdot)+|V|(\cdot)\right)f\rangle_{\ell^2(\Vc,m)},
\end{align*}
with
\[\deg_1(x):=\frac{1}{m(x)}\sum_{y\in \Vc} \Ec(x,y)\frac{|\xi(x,y)|}{\sqrt{(1+\mu(y))(1+\mu(x))}},\]
\begin{align*}
\deg_2(x):=
\frac{1}{m(x)}\sum_{y\in \Vc} \Ec(x,y)\left(\frac{|\mu(x)+\mu(y)+\mu(x)\mu(y)|}{\sqrt{(1+\mu(y))(1+\mu(x))}
(1+\sqrt{(1+\mu(y))(1+\mu(x))})}\right),
\end{align*}
and
\[\deg_3(x):=\frac{1}{m(x)}\sum_{y\in \Vc} \Ec(x,y)\left|e^{\rmi \theta(x,y)}-\frac{e^{\rmi \theta_\gamma(x,y)}}{\sqrt{(1+\mu(y))(1+\mu(x))}}\right|.\]
Finally, by $(H_0)$, $|V|(x)$, $|W|(x)$, $\deg_1(x),$ $\deg_2(x),$ and $\deg_3(x)$ are $o((1+\deg(x)))$, as
$|x|\to \infty$, by density of $\Cc_c(\Vc)$, it guarantees:
\begin{align}\label{e:compdom}
(\deg_{\Gc}(\cdot)+1)^{-1/2}( \Delta_{\Gc^{}}- H)(\deg_{\Gc}(\cdot)+1)^{-1/2} \in \Kc(\ell^2(\Vc, m).
\end{align}
In particular, $ \Delta_{\Gc^{}}-  H$ is a
infinitesimally small form perturbation of $\Delta_{\Gc}$. By the KLMN Theorem, e.g.\ \cite[Theorem X.17]{RS}, we infer that $\Gr:=\Dc(\Delta_{\Gc}^{1/2}) =
\Dc(| H|^{1/2})$.
It remains to use the resolvent identity. Formally, we have:
\begin{align}\label{e:resoform}
( H+\rmi)^{-1}-(\Delta_{\Gc^{}}+\rmi)^{-1}
= ( H+\rmi)^{-1}
( \Delta_{\Gc^{}}-  H)
(\Delta_{\Gc^{}}+\rmi)^{-1}.
\end{align}
However,  as we have to work
with forms,  one should be careful with the resolvent equation. We
give a complete proof and refer to \cite{GG2} for more discussions of
this matter. To lighten notation, we set $H_0:=\Delta_{\Gc}$ and $H:=\Delta_{\Gc_{\varepsilon, \mu,\gamma}}+V(\cdot)$.
To start off, both operators
extend to an element of $\Bc(\Gr, \Gr^*)$, by \eqref{e:inclusion}.  We denote these
extensions with a tilde.
We have $(H_0+\rmi)^{-1*}\Hc\subset\Gr$. This allows one to
deduce that ($H_0+\rmi)^{-1}$ extends to a unique continuous operator
$\Gr^*\rightarrow\Hc$. We denote it for the moment by $R$. From
$R(H_0+\rmi)u=u$ for $u\in \Dc(H_0)$ we get, by density of
$\Dc(H_0)$ in $\Gr$ and continuity, $R(\widetilde{H_0}+\rmi)u=u$ for
$u\in\Gr$. In particular
\begin{equation*}
(H+\rmi)^{-1}= R(\widetilde{H_0}+\rmi)(H+\rmi)^{-1}.
\end{equation*}
Clearly,
\begin{equation*}
(H_0+\rmi)^{-1}= (H_0+\rmi)^{-1}(H+\rmi)(H+\rmi)^{-1}=
R(\widetilde{H}+\rmi)(H+\rmi)^{-1}.
\end{equation*}
We subtract the last two relations to obtain that
\begin{equation*}
(H_0+\rmi)^{-1}-(H+\rmi)^{-1}=R(\widetilde{H}
-\widetilde{H_0})(H+\rmi)^{-1}.
\end{equation*}
Since $R$ is uniquely determined as the extension of $(H_0+\rmi)^{-1}$ to a
continuous map $\Gr^*\rightarrow\Hc$, one may keep the notation
$(H_0+\rmi)^{-1}$ for it. With this convention, the rigorous version of
(\ref{e:resoform}) that we use is:
\begin{equation}\label{e:resoform2}
(H_0+\rmi)^{-1}-(H+\rmi)^{-1}=(H_0+\rmi)^{-1}(\widetilde
H-\widetilde H_0)(H+\rmi)^{-1}.
\end{equation}
Recalling that $\Gr= \Dc(\deg_\Gc^{1/2}(\cdot))$, \eqref{e:compdom} ensures that
$\widetilde
H-\widetilde H_0$ is a compact operator from $\Gr$ to $\Gr^*$. In particular,
$(H_0+\rmi)^{-1}-(H+\rmi)^{-1}$
is in $\Kc(\ell^2(\Vc, m)$ and the Weyl's Theorem gives the essential
spectrum.
\qed

\subsection{Commutators of the perturbation}
We start with a technical lemma so as to apply \cite[Proposition 7.5.7]{ABG}.
To deal with a large class of perturbation, the $\Cc^2(\Ac_\Gc, \Gr, \Gr^*)$ theory will not be enough and we introduce subclasses.\\
In the abstract setting of Section \ref{S,f}:
 \begin{enumerate}[label=\roman*.]
\item We say that  $H\in\Cc^{0,1}(\Ac,\Gr,\Gr^*)$,  if $\int^1_0 \left\| [(H+\rm i)^{-1},e^{\rm i t\Ac}]\right\|_{B(\Gr,\Gr^*)} \frac{dt}{t}<\infty$.
\item We say that  $H\in\Cc^{1,1}(\Ac,\Gr,\Gr^*)$,  if $\int^1_0 \left\| [[(H+\rm i)^{-1},e^{\rm i t\Ac}],e^{\rm i t\Ac}]\right\|_{B(\Gr,\Gr^*)} \frac{dt}{t^2}<\infty$.
\end{enumerate}
Thanks to \cite[p. 205]{ABG}, it turns out that
\[\Cc^2(\Ac,\Gr,\Gr^{*})\subset\Cc^{1,1}(\Ac,\Gr,\Gr^{*})\subset\Cc^{1,u}(\Ac,\Gr,\Gr^{*})\subset\Cc^1(\Ac,\Gr,\Gr^{*})\subset \Cc^{0,1}(\Ac,\Gr,\Gr^{*}).\] The next result aims at replacing partially $\Ac_\Gc$ to prove the $\Cc^{1,1}(\Ac_\Gc,\Gr,\Gr^{*})$ criteria.
\begin{proposition}\label{p:hypothesis}
Let $\Lambda:=\langle Q\rangle \otimes \un_{\Vc_2}$.
It extends by density to $\Lambda|_{\Gr^*}$. We denote its extension by $\Lambda|_{\Gr^*}$.
 We have:
\begin{enumerate}
\item $\Dc(\Lambda|_{\Gr^*})\subset \Dc(\Ac_{\Gr^*})$.
\item There is $c>0$ such that for all $r>0$, $-\rmi r$ belongs to the resolvent set of $\Lambda|_{\Gr^*}$ and $r\|(\Lambda|_{\Gr^*}) +\rmi r)^{-1}\|_{\Bc(\Gr^*)}\leq c$.
\item $t\mapsto e^{\rmi t\Lambda|_{\Gr^*}}$  has a polynomial growth in $\Gc^*$.
\item Given $\xi\in \Cc^\infty(\R;\R)$ such that $\xi(x)=0$ near $0$ and $1$ near infinity and $T\in \Bc(\Gr, \Gr^*)$ symmetric, if
\begin{align}
\int_1^\infty \left\| \xi\left(\frac{\Lambda|_{\Gr^*}}{r}\right)T\right\|_{\Bc(\Gr, \Gr^*)} \, dt <\infty
\end{align}
then $T\in \Cc^{0,1}(\Ac_\Gc; \Gr, \Gr^*)$.
\end{enumerate}
\end{proposition}
\proof (1) Recalling the action of $T$ and the trivial action
of the operator on the second part of the tensor product,
it is enough to prove $\Dc(\langle Q\rangle)\subset \Dc(\Ac_\N)$ in $\ell^2(\N,1)$. This follows directly from \eqref{e:AN}.

(2) Note that $\Lambda|_{\Gr^*}$ is self-adjoint in $\Gr^*$,
thanks to its trivial action of the second part of the tensor product.
The result is therefore trivial.

(3) Again, since $\Lambda$ is self-adjoint in $\Gr^*$, the norm of $t\mapsto e^{\rmi t\Lambda|_{\Gr^*}}$ is $1$ for all $t\in \R$. It has in particular polynomial growth.

(4) Apply \cite[Proposition 7.5.7]{ABG}. \qed

This abstract result will be replaced with the following simple criteria.
\begin{corollary}\label{c:C01}
With the notation of Proposition \ref{p:hypothesis}, let $\varepsilon \in (0,1)$ and $T\in \Bc(\Gr, \Gr^*)$ symmetric.
Assume that:
\[
\langle\Lambda|_{\Gr^*}\rangle^{\varepsilon}\, T  \in \Bc(\Gr, \Gr^*),
\]
then $T\in \Cc^{0,1}(\Ac_\Gc; \Gr, \Gr^*)$.
\end{corollary}
We shall also need to replace $\Lambda$ is the Limiting Absorption Principle given with weights in $\Ac_\Gc$.
This is guaranteed by the next remark.
\begin{lemma}\label{l:interpo}
Given $s\in [0,1
]$, there is $c_s\geq 0$ such that
\begin{align}\label{e:interpo}
 \|\langle \Ac_{\overline{\Gc},\Gr}\rangle^s f\|_{\Gr}&\leq c_s \|(\langle Q\rangle\otimes \id_{\Dc(1/m_{2}^{1/2})} )^s f\|_{\Gr} = c_s \|(\langle Q\rangle^s\otimes \id_{\Dc(1/m_{2}^{1/2})}) f\|_{\Gr}\\&\leq c_s\|( \Lambda|_{\Gr})^s f\|^2_{\Gr}\nonumber,\end{align}  for all  $f\in \Cc_c(\Vc)$.
\begin{align}\label{e:interpo2}
 \|\langle \Ac_{\overline{\Gc},\Gr^{*}}\rangle^s f\|_{\Gr^{*}}&\leq c_s \|(\langle Q\rangle\otimes \id_{\Dc(m_{2}^{1/2})} )^s f\|_{\Gr^{*}} = c_s \|(\langle Q\rangle^s\otimes \id_{\Dc(m_{2}^{1/2})}) f\|_{\Gr^{*}}\\&\leq c_s\| (\Lambda|_{\Gr^*})^s f\|^2_{\Gr^{*}}\nonumber,\end{align} for all $f\in \Cc_c(\Vc)$.

\end{lemma}
\proof\ We will prove only  \ref{e:interpo}. Note that  \ref{e:interpo2}
is established in very same  manner.\\ By Riesz-Thorin interpolation theory, it is enough to prove it for $s=1$.
 So, by using \eqref{dec} and Remark \ref{r:geneG}, there are $C_0$ and $C_1$ bounded in $\ell^2(\Vc_1, m_1)$, so that
for all $f\in\Cc_c(\Vc)$,
\begin{align*}
\|\Ac_{\overline{\Gc}}f\|^2_{\Gr} &=\|\left( C_1Q \otimes \id_{\Dc(1/m_{2}^{1/2})} +
C_0\right)\otimes \id_{\Dc(1/m_{2}^{1/2})} f\|^2 _{\Gr}\leq c\| \Lambda|_{\Gr} f\|^2_{\Gr}.
\end{align*}
By real interpolation, e.g \cite[Theorem 4.1.2,p.88]{BL}, for all $s\in[0,1]$
there is $c_s\geq0$ such that
\begin{align*}\|\langle \Ac_{\overline{\Gc},\Gr}\rangle^s f\|_{\Gr}\leq c_s \|(\langle Q\rangle\otimes \id_{\Dc(1/m_{2}^{1/2})} )^s f\|_{\Gr}
&=c_s \|(\langle Q\rangle^s\otimes \id_{\Dc(1/m_{2}^{1/2})} ) f\|_{\Gr} &\\ \leq c\| (\Lambda|_{\Gr})^{s} f\|^2_{\Gr},\nonumber \end{align*} for all  $f\in \Cc_c(\Vc)$.
Finally, we  conclude by density.\qed
\begin{remark} As $\Lambda$ commutes with $T_{m_{\mu}\rightarrow m}$, the Proposition \ref{p:hypothesis}, Corollary \ref{c:C01},
 and Lemma \ref{l:interpo} hold true with $\Ac_{\Gc_{\xi, \mu,\gamma}}$.
In $\ell^2(\Vc, m_\mu)$, we have:
\begin{align*}
\Ac_{\Gc_{\xi, \mu,\gamma}}:=-\frac{\rmi}{2\sqrt{1+\mu(\cdot)}}&\Big(e^{1/2}(1/2+Q)e^{-\rmi \theta_{Q,Q+1}}U^\ast\sqrt{1+\mu(\cdot)}
\\
&+e^{-1/2}(1/2-Q)e^{-\rmi \theta_{Q,Q-1}}U\sqrt{1+\mu(\cdot)}\Big)\otimes\un_{\Vc_2}.
\end{align*}
\end{remark}
\subsection{Regularity of the potential}
For the sake of the reader, we have separated the treatment of the potential $V$ to present the technical steps.
\begin{lemma}\label{l:Vfun}Let $V:\Vc \to\R$ be a function. We assume that $(H_4)$ holds true,
then $V(\cdot)\in \Cc^1(\Ac_{\Gc_{\xi, \mu,\gamma}}; \Gr, \Gr^*)$ and
 $[V(\cdot),\Ac_{\Gc_{\xi, \mu,\gamma}}]_\circ\in\Cc^{0,1}(\Ac_{\Gc_{\xi, \mu,\gamma}}; \Gr, \Gr^*)$. In particular, $V(\cdot)\in\Cc^{1,1}(\Ac_{\Gc_{\xi, \mu,\gamma}} ; \Gr, \Gr^*)$.
\end{lemma}
\proof As $T_{m_\mu\to m}$ commutes with $V$,  it is enough to deal with $\xi=\mu=\gamma=0$.
We start with the $\Cc^1$ property, as in Remark \ref{r:comH}. We compute on $\Cc_c(\Vc)$ and aim at boundedness in $\Bc(\Gr, \Gr^*)$. Recalling Proposition \ref{p:AG}, we have:
\begin{align*}
2[V(Q), \rmi \Ac_\Gc] &=  2\Re([V(Q), \rmi S_\Gc Q \otimes \un_{\Vc_2}]) = \Re([V(Q),  (U_\Gc-U_\Gc^*) Q \otimes \un_{\Vc_2}])
\end{align*}
Therefore we need to bound four terms:
\begin{align*}
[V(Q),  U_\Gc\otimes \un_{\Vc_2}] Q \otimes \un_{\Vc_2}\quad &; \quad [V(Q),  U_\Gc^*\otimes \un_{\Vc_2}] Q \otimes \un_{\Vc_2}
\\
Q \otimes \un_{\Vc_2} [V(Q),  U_\Gc\otimes \un_{\Vc_2}] \quad &; \quad
Q \otimes \un_{\Vc_2}[V(Q),  U_\Gc^*\otimes \un_{\Vc_2}].
\end{align*}
By exploiting the adjoint property, it is sufficient to control the first two commutators. Recalling that $U_{\Gc}:= e^{\rm i \theta_1(Q,Q-1)} e^{-1/2}U$,
\begin{align*}
[V(Q), U_\Gc Q\otimes \un_{\Vc_2}] f(n_1, n_2)&=
\\
&\hspace*{-3cm}
e^{-1/2}e^{\rm i \theta_1(n_1,n_1-1)} (n_1-1)(V(n_1, n_2)- V(n_1-1, n_2)) Uf(n_1, n_2).
\end{align*}
It is bounded in $\Bc(\Gr, \Gr^*)$ by $(H_4)$. Then, since
$U_{\Gc}^*= e^{-\rm i \theta_1(Q+1,Q)} e^{1/2}U^*$,
\begin{align*}
[V(Q), U_\Gc^* Q\otimes \un_{\Vc_2}] f(n_1, n_2)&=
\\
&\hspace*{-3cm}
e^{1/2}e^{\rm i \theta_1(n_1+1,n_1)} (n_1+1)(V(n_1, n_2)- V(n_1+1, n_2)) U^*f(n_1, n_2)
\end{align*}
It also belongs to $\Bc(\Gr, \Gr^*)$ by $(H_4)$. We conclude that $V(\cdot)\in \Cc^1(\Ac_{\Gc_{\xi, \mu,\gamma}}; \Gr, \Gr^*)$.

We turn to the $\Cc^{1,1}$ property. Recalling that $\epsilon\in (0,1)$. We denote by $c$ a generic constant
which is independent of $f\in \Cc_c(\Vc)$, we have:
\begin{align*}
\|\langle \Lambda|_{\Gr^*} \rangle^\epsilon [V(Q), \Ac_\Gc]_\circ f\|_{\Gr^*} &
\\
&\hspace*{-3.5cm}
\leq
c \|\langle Q_1\rangle^\epsilon (V(Q_1, Q_2)- V(Q_1-1, Q_2)) U\otimes \un_{\Vc_2} f \|_{\Gr^*}
\\
&\hspace*{-3.5cm}\,\,  + c \|\langle Q_1\rangle^\epsilon (V(Q_1, Q_2)- V(Q_1-1, Q_2)) U^*\otimes \un_{\Vc_2}f \|_{\Gr^*}
\\
&\hspace*{-3.5cm}\leq c \|\langle Q_1\rangle^\epsilon (V(Q_1, Q_2)- V(Q_1-1, Q_2)) U\otimes \un_{\Vc_2} (1+\deg_{\Gc}(Q))\|_{\Bc(\Gr, \Gr^*)} \cdot  \| f \|_{\Gr}
\\
&\hspace*{-3.5cm} \,\, + c \|\langle Q_1\rangle^\epsilon (V(Q_1, Q_2)- V(Q_1-1, Q_2)) U^*\otimes \un_{\Vc_2} (1+\deg_{\Gc}(Q))\|_{\Bc(\Gr, \Gr^*)} \cdot  \| f \|_{\Gr}
\\
& \hspace*{-3.5cm} \leq c \|f|\|_{\Gr},
\end{align*}
where we have used $(H_4)$ and the commutation with $U$ and $U^*$. Conclude by density of
$\Cc_c(\Vc)$ in $\Gr$ and Corollary \ref{c:C01}. \qed
\subsection{Regularity of the metric}
We now give a criteria to ensure that an operator is in $\Bc(\Gr, \Gr^*)$.

\begin{lemma}\label{l:critbd}
Let $B:\Vc\times \Vc \to \C$. For all $f\in \Cc_c(\Vc)$, set
\[Tf(x):= \sum_{y} B(x,y) f(y). \]
Assume
\[M:= \sup_{x,y \in \Vc} \frac{m(x) |B(x,y)|+ m(y)|B(x,y)|}{\Ec(x,y)}<\infty.\]
Then, for $f\in \Cc_c(\Vc)$, we have:
\[|\langle f, Tf \rangle| \leq M \langle f, \deg_{\Gc}(\cdot)f \rangle \]
and in particular, (the closure of) $T\in \Bc(\Gr, \Gr^*)$.
\end{lemma}
\proof
Let $f\in \Cc_c(\Vc)$.
\begin{align*}
|\langle f, T f \rangle| &= |\sum_{x\in \Vc} m_\Gc (x)  \overline{f(x)} \sum_{y\in \Vc} B(x,y) f(y) |
\\
&\leq \frac{1}{2} \sum_{x, y \in \Vc} m_\Gc (x)\, |B(x,y)|\, ( |f(x)|^2+ |f(y)|^2) \leq M \langle f, \deg_{\Gc}(\cdot) f\rangle.
\end{align*}
Conclude by density of $\Cc_c(\Vc)$ in $\Gr$. \qed

Now, we introduce  a slightly modified version of \cite[Lemma 3.27]{AEGJ2} adapted to our context.
\begin{proposition}\label{P:sym}Let $\omega$ be a nonnegative function defined on $\Z\times\Vc_2$ and $P:\Z\times\Vc_2\longrightarrow\C$, for which
\begin{align}\label{lim0}\lim_{|(x_1,x_2)|\rightarrow\infty}P(x_1,x_2)=0,\end{align}
and a fixed $\varepsilon  >0$ satisfies \begin{align}\label{h:sym}\sup_{(x_1,x_2)\in\Z\times\Vc_2}\omega(x)\Lambda^{\varepsilon +1}(x)\Big|P(x_1-1,x_2)-P(x_1,x_2)\Big|<\infty.
\end{align}
Then
\begin{align*}\sup_{(x_1,x_2)\in\Z\times\Vc_2}\omega(x)\Lambda^{\varepsilon }(x)\Big|P(x_1,x_2)\Big|<\infty.
\end{align*}
\end{proposition}
\proof
Let $x=(x_1,x_2),y=(y_1,x_2)\in\Z\times\Vc_2$. We denote any constant by $M$, which is independent of $x$.\\
First, we treat the case $x_1>0$, and we assume that $y_1>x_1+1$.

\begin{align*}&\omega(x)\Lambda^\varepsilon (x)\Big|P(y_1,x_2)-P(x_1,x_2)\Big|
\leq\omega(x)\Lambda^\varepsilon (x)\sum_{k=0}^{y_1-x_1-1}\Big|P(x_1+k+1,x_2)-P(x_1+k,x_2)\Big|
\\
&\hspace*{-0.3cm}\leq\omega(x)\Lambda^\varepsilon (x)\sum_{k=0}^{y_1-x_1-1}\langle x_1+k\rangle^{\varepsilon +1}
\langle x_1+k\rangle^{-\varepsilon -1}\Big|P(x_1+k+1,x_2)-P(x_1+k,x_2)\Big|.\end{align*}
By using \eqref{h:sym}, we get
\begin{align*}&\omega(x)\Lambda^\varepsilon (x)\Big|P(y_1,x_2)-P(x_1,x_2)\Big|
\leq M\Lambda^\varepsilon (x)\underbrace{\sum_{k=0}^{y_1-x_1-1}\langle x_1+k\rangle^{-1-\varepsilon }}_{=:I}.
\end{align*}
We leverage that, $k\mapsto\langle x_1+k\rangle^{-1-\varepsilon }$ is a decreasing function on $[0,+\infty[$,
and by using
\begin{align}\label{ing2}\langle t\rangle^{-1}\leq\frac{M}{|t|+1}, \hbox{ for all }t\in\R.
\end{align}
We get,
\begin{align*}I&\leq M\int_{0}^{y_1-x_1}\frac{1}{(x_1+t+1)^{1+\varepsilon }}dt=\frac{M}{\varepsilon }
\Bigg(\frac{1}{(x_1+1)^\varepsilon }-\frac{1}{(y_1+1)^\varepsilon }\Bigg).
\end{align*}
If $x_1\leq0$, we take $y_1<x_1-1$. We obtain
\begin{align*}&\omega(x)\Lambda^\varepsilon (x)\Big|P(y_1,x_2)-P(x_1,x_2)\Big|
\leq\omega(x)\Lambda^\varepsilon (x)\sum_{k=0}^{x_1-y_1-1}\Big|P(x_1-k-1,x_2)-P(x_1-k,x_2)\Big|
\\
&\hspace*{-0.3cm}\leq\omega(x)\Lambda^\varepsilon (x)\sum_{k=0}^{x_1-y_1-1}\langle x_1-k\rangle^{\varepsilon +1}
\langle x_1-k\rangle^{-\varepsilon -1}\Big|P(x_1-k-1,x_2)-P(x_1-k,x_2)\Big|
\\
&\hspace*{-0.3cm}\leq M\Lambda^\varepsilon (x)\underbrace{\sum_{k=0}^{x_1-y_1-1}
\langle x_1-k\rangle^{-1-\varepsilon }}_{=:J}.
\end{align*}
Since $k\mapsto\langle x_1-k\rangle^{-1-\varepsilon }$ is a increasing function on $]-\infty,0]$,
and by using \eqref{ing2}.
We have:
\begin{align*}J&\leq\frac{M}{\varepsilon }\Bigg(\frac{1}{(-y_1+1)^\varepsilon }-\frac{1}{(-x_1+1)^\varepsilon }\Bigg).
\end{align*}
Therefore,
\begin{align*}\omega(x)\Lambda^{\varepsilon }(x)\Big|P(x_1,x_2)\Big|
\leq\frac{M}{\varepsilon }\Bigg(\frac{1}{(|y_1|+1)^\varepsilon }+\frac{1}{(|x_1|+1)^\varepsilon }\Bigg).
\end{align*}
We tend $|y_1|$ to $+\infty$, and through the use of \eqref{lim0},
we conclude that: $$\omega(x)\Lambda^{\varepsilon }(x)\Big|P(x_1,x_2)\Big|\leq M.$$ \qed

Also, it is necessary to provide the antisymmetric variant  of the precedent proposition.
\begin{proposition}\label{P:anti}Consider a nonnegative function $\omega$ defined on $\Z\times\Vc_2$. Let $\varphi:(\Z\times\Vc_2)^2\longrightarrow\C$, an antisymmetric function, for which
\begin{align}\label{anti}\lim_{|(x_1,x_2)|,|(y_1,y_2)|\rightarrow\infty}\varphi((x_1,x_2),(y_1,y_2))=0,\end{align}
and fixed $\varepsilon  >0$ satisfies
\begin{align}\label{h:anti}\sup_{(x_1,x_2)\in\Z\times\Vc_2}\omega(x)\Lambda^{\varepsilon+1}(x)
\Big|\varphi((x_1,x_2),(x_1+1,x_2))+\varphi((x_1-1,x_2),(x_1,x_2))\Big|<\infty,
\end{align}
then
\begin{align*}\sup_{(x_1,x_2)\in\Z\times\Vc_2}\omega(x)\Lambda^{\varepsilon}(x)\Big|\varphi((x_1,x_2),(x_1+1,x_2))\Big|<\infty.
\end{align*}
\end{proposition}
\proof
Let $x=(x_1,x_2),y=(y_1,x_2)\in\Z\times\Vc_2$. We denote any constant by $M$, which is independent of $x$.
First, we treat the case $x_1>0$, and we assume that $y_1>x_1+1$.
If $y_1-x_1$ is even.
\begin{align*}&\omega(x)\Lambda^\varepsilon (x)\Big|\varphi((x_1,x_2),(x_1+1,x_2))+\varphi((y_1+1,x_2),(y_1+2,x_2))\Big|
\\
&\hspace*{-0.3cm}\leq\omega(x)\Lambda^\varepsilon (x)\sum_{k=0}^{y_1-x_1}\langle x_1+k\rangle^{\varepsilon +1}
\langle x_1+k\rangle^{-1-\varepsilon }
\\
&\quad \times\Big|\varphi((x_1+k,x_2),(x_1+k+1,x_2))
+\varphi((x_1+k+1,x_2),(x_1+k+2,x_2))\Big|
\\
&\hspace*{-0.3cm}\leq M\Lambda^\varepsilon (x)\sum_{k=0}^{y_1-x_1}\langle x_1+k\rangle^{-1-\varepsilon }.
\end{align*}
We conclude as in Proposition \ref{P:sym}. In same way, we treat the case, $y_1-x_1$ is odd.
\begin{align*}&\omega(x)\Lambda^\varepsilon (x)\Big|\varphi((x_1,x_2),(x_1+1,x_2))+\varphi((y_1,x_2),(y_1+2,x_2))\Big|
\\
&\hspace*{-0.5cm}=\omega(x)\Lambda^\varepsilon (x)\Bigg|\sum_{k=0}^{y_1-x_1-1}(-1)^k\Big(\varphi((x_1+k,x_2),(x_1+k+1,x_2))
\\
&\quad+\varphi((x_1+k+1,x_2),(x_1+k+2,x_2))\Big)\Bigg|.
\end{align*}
In the same way, we deal with the case $x_1\leq0$.
Therefore,
\begin{align*}\omega(x)\Lambda^{\varepsilon }(x)&\Big|\varphi((x_1,x_2),(x_1+1,x_2))\Big|
\leq\frac{M}{\varepsilon }\Bigg(\frac{1}{(|y_1|+1)^\varepsilon }+\frac{1}{(|x_1|+1)^\varepsilon }\Bigg).
\end{align*}
We tend $|y_1|$ to $+\infty$, and through the use of \eqref{lim0},
we conclude that
\[\omega(x)\Lambda^{\varepsilon }(x)\Big|\varphi((x_1,x_2)(x_1+1,x_2))\Big|\leq M.\] \qed
\begin{proposition}\label{C1,1fun}
Assuming $(H1)$, $(H2)$, and $(H3)$ hold true,
we have: $${\Delta}_{\Gc_{\xi, \mu, \gamma}}\in \Cc^1(\Ac_{\Gc_{\xi, \mu,\gamma}}; \Gr, \Gr^*).$$ Moreover
$[{\Delta}_{\Gc_{\xi, \mu, \gamma}},\Ac_{\Gc_{\xi, \mu,\gamma}}]_\circ\in\Cc^{0,1}(\Ac_{\Gc_{\xi, \mu,\gamma}}; \Gr, \Gr^*)$. In particular, we have that: $${\Delta}_{\Gc_{\xi, \mu, \gamma}}\in\Cc^{1,1}(\Ac_{\Gc_{\xi, \mu,\gamma} }; \Gr, \Gr^*).$$
\end{proposition}
\proof Recalling Lemma \ref{l:C2Af} and Remark \ref{r:transfert}, it is enough to solely prove that:
$$\widetilde\Delta_{\Gc_{\xi, \mu, \gamma}^{}}- \Delta_\Gc\in\Cc^{1}(\Ac_{\Gc}; \Gr, \Gr^*) \mbox{ and }
[\widetilde\Delta_{\Gc_{\xi, \mu, \gamma}^{}}- \Delta_\Gc,\Ac_{\Gc}]_\circ\in\Cc^{0,1}(\Ac_\Gc; \Gr, \Gr^*).$$
In the first step, we are going to show  that $\widetilde\Delta_{\Gc_{\xi, \mu, \gamma}^{}}\in \Cc^{1}(A,\Gr,\Gr^{*})$. It suffices to show that there exists $c>0$, such that:

\begin{equation}\label{eq1}
\left\|\left[\widetilde\Delta_{\Gc_{\xi, \mu, \gamma}^{}}- \Delta_\Gc, \rmi \Ac\right]f\right\|_{\Gr^{*}}^2\leq c \|f\|_{\Gr}^2,\ \forall f\in \Cc_{c}(\Vc).
\end{equation}
Then, by density and thanks to \cite[Proposition 6.2.9]{ABG} and Lemma \ref{l:C2Af}, we obtain the result.
In the second step, we will establish that $[\widetilde\Delta_{\Gc_{\xi, \mu, \gamma}^{}},\rmi \Ac]_{\circ}\in\Cc^{0,1}(\Ac_\Gc; \Gr, \Gr^*)$. Given $\varepsilon\in[0,\epsilon)$, where $\epsilon\in (0,1)$  was fixed in the hypotheses. We show there
exists $c_{\varepsilon} > 0$ such that:
\begin{equation}\label{eq2}
\left\|(\Lambda|_{\Gr^{*}})^{\varepsilon}\left[\widetilde\Delta_{\Gc_{\xi, \mu, \gamma}^{}}- \Delta_\Gc,\Ac_{\Gc},\rmi \Ac\right]f\right\|_{\Gr^{*}}^2\leq c_{\varepsilon}\|f\|_{\Gr}^2,\ \forall F \Cc_{c}(\Vc).
\end{equation}
Then, by density, $[\widetilde\Delta_{\Gc_{\xi, \mu, \gamma}^{}}- \Delta_\Gc,\Ac_{\Gc},\rmi \Ac]_{\circ}\in \Cc^{0,1}(A,\Gr,\Gr^{*})
$. Finally, by
Corollary \ref{c:C01} and thanks to Lemma \ref{l:C2Af}, we get $[\widetilde\Delta_{\Gc_{\xi, \mu, \gamma}^{}},\rmi \Ac]_{\circ}\in \Cc^{0,1}(\Ac_\Gc; \Gr, \Gr^*)$. In particular,  by Lemma \ref{l:C2Af}, we obtain $\widetilde\Delta_{\Gc_{\xi, \mu, \gamma}^{}}\in \Cc^{1,1}(\Ac_\Gc; \Gr, \Gr^*).$\\
In \eqref{e:diff}, we have four terms. They share some similarities. We focus on the third one. Let $f\in \Cc_c(\Vc)$. We write
\[Df(x) = \sum_{y\in \Vc} D(x,y) f(y), \]
with
\begin{align*}
D(x,y)&:=\frac{1}{m^{}(x)} \Ec^{}(x,y)\left(e^{\rmi \theta(x,y)}-\frac{e^{\rmi \theta_\gamma(x,y)}}{\sqrt{(1+\mu(y))(1+\mu(x))}}\right)
\\
&=\frac{1}{m(x)}\Ec(x,y)\frac{e^{\rmi\theta(x,y)}}{\sqrt{(1+\mu(x))(1+\mu(y))}}
\Big(\frac{\mu(x)+\mu(y)+\mu(x)\mu(y)}{\sqrt{(1+\mu(x))(1+\mu(y))}+1}\Big)
\\
&\quad+\frac{1}{m(x)}\Ec(x,y)\frac{e^{\rmi\theta(x,y)}}{\sqrt{(1+\mu(x))(1+\mu(y))}}\Big(1-e^{\rmi\gamma(x,y)}\Big)
\\
&=D_1(x,y)+D_2(x,y).
\end{align*}
To lighten the notation, given $x=(x_1, x_2)\in \Vc$, we write $x\pm1$ for $(x_1\pm 1, x_2)$. Note that
\begin{align*}
[D, U^*] f (x) &= \sum_{y\in \Vc} (D(x,y)- D(x+1, y)) (U^*f)(y)
\\
&= \sum_{y\in \Vc} (D(x,y-1)- D(x+1, y-1))\un_{[1, \infty[}(y) f(y).
\end{align*}
Recalling Proposition \ref{p:AG}, we start by showing the boundedness in $\Bc(\Gr, \Gr^*)$ of
$$\langle Q_1\rangle^{\varepsilon} [D, Qe^{\rm i \theta_1(Q-1,Q)}U^*].$$
A straightforward computation leads to
\begin{align}
\nonumber
[D_2,Q_1U^*]f(x)&=\frac{1}{m(x)}x_1\Ec_1(x_1,x_1+1)
\frac{1-e^{\rmi \gamma(x,x+1)}}{\sqrt{(1+\mu(x))(1+\mu(x+1))}}f(x+2)
\\
\nonumber
&\hspace*{-1.25cm}-\frac{1}{m(x+1)}x_1\Ec_1(x_1+1,x_1+2)
\frac{1-e^{\rmi \gamma(x+1,x+2)}}{\sqrt{(1+\mu(x+1))(1+\mu(x+2))}}f(x+2)
\\
&\label{H:7}+\frac{1}{m(x)}\Ec_1(x_1,x_1+1)
\frac{1-e^{\rmi \gamma(x,x+1)}}{\sqrt{(1+\mu(x))(1+\mu(x+1))}}f(x+2)
\\
\nonumber
&+\frac{1}{m(x)}x_1\Ec_1(x_1,x_1-1)
\frac{1-e^{\rmi \gamma(x,x-1)}}{\sqrt{(1+\mu(x))(1+\mu(x-1))}}f(x)
\\
\nonumber
&-\frac{1}{m(x+1)}x_1\Ec_1(x_1,x_1+1)
\frac{1-e^{\rmi \gamma(x+1,x)}}{\sqrt{(1+\mu(x))(1+\mu(x+1))}}f(x)
\\
\label{e:1}
&-\frac{1}{m(x+1)}\Ec_1(x_1,x_1-1)
\frac{1-e^{\rmi \gamma(x-1,x)}}{\sqrt{(1+\mu(x))(1+\mu(x-1))}}f(x)
\\
\nonumber
&+x_1\sum_{y_2\in\Vc_2}\Bigg(\frac{1}{m(x)}\Ec_2(x_2,y_2)
\frac{(1-e^{\rmi \gamma(x,(x_1,y_2))})e^{\rmi \theta_2(x_2,y_2)}}{\sqrt{(1+\mu(x))(1+\mu(x_1,y_2))}}
\\
\label{e:2}
\\
\nonumber
&-\frac{1}{m(x+1)}\Ec_2(x_2,y_2)
\frac{(1-e^{\rmi \gamma(x+1,(x_1+1,y_2))})e^{\rmi \theta_2(x_2,y_2)}}{\sqrt{(1+\mu(x+1))(1+\mu(x_1+1,y_2))}}\Bigg)\\
\nonumber
&\times\un_{[1,+\infty[}(x_1+1)f(x_1+1,y_2).
\end{align}
To begin with, we carefully examine the last term. To simplify we denote it by $B$ the operator expressed as follows:  \[Bf(x):=x_1\sum_{y_2\in\Vc_2}\frac{1}{m(x)}\Ec_2(x_2,y_2)
\frac{(1-e^{\rmi \gamma(x,(x_1,y_2))})e^{\rmi \theta_2(x_2,y_2)}}{\sqrt{(1+\mu(x))(1+\mu(x_1,y_2))}}\un_{[1,+\infty[}(x_1+1)f(x_1+1,y_2).\]

\begin{align*}\Big\|(\Lambda|_{\Gr^*})^{\varepsilon}(\cdot)Bf\Big\|_{\Gr^*}^2
&=\Big\|\Big(1+\deg_{\Gc}(\cdot)\Big)^{-\frac{1}{2}}(\Lambda|_{\Gr^*})^{\varepsilon}(\cdot)Bf\Big\|^2\\&=
\sum_{x\in\Vc}m(x)\Bigg|\Big(1+\deg_{\Gc}(x)\Big)^{-\frac{1}{2}}(\Lambda|_{\Gr^*})^{\varepsilon}(x)x_1
\sum_{y_2\in\Vc_2}\frac{\Ec_2(x_2,y_2)}{m(x)}e^{\rmi\theta_2(x_2,y_2)}
\\
&\quad\times\frac{1-e^{\rmi \gamma(x,(x_1,y_2))}}{\sqrt{(1+\mu(x))(1+\mu(x_1,y_2))}}f(x_1+1,y_2)\Bigg|^2
\\
&\leq8\sum_{x_1\in\Vc_1}m_1(x_1)\frac{(\Lambda|_{\Gr^*})^{2\varepsilon}(x)x_1^2}{(m_1(x_1))^2}\\&\sum_{x_2\in\Vc_2}m_2(x_2)\Big(1+\deg_{\Gc}(x_1,x_2)\Big)^{-1}
\sum_{y_2\in\Vc_2}\Big(\frac{\Ec_2(x_2,y_2)}{m_2(x_2)}\Big)^2\\
&\quad\times\frac{\sin^2(\frac{\gamma(x,(x_1,y_2))}{2})}{(1+\mu(x))(1+\mu(x_1,y_2))}\Big|f(x_1+1,y_2)\Big|^2.
\end{align*}
By invoking Fubini's theorem, we have
\begin{align*}\Big\|(\Lambda|_{\Gr^*})^{\varepsilon}(\cdot)Bf\Big\|_{\Gr^*}^2&=
\Big\|\Big(1+\deg_{\Gc}(\cdot)\Big)^{-\frac{1}{2}}(\Lambda|_{\Gr^*})^{\varepsilon}(\cdot)Bf\Big\|^2\\&\leq8\sum_{x_1\in\Vc_1}m_1(x_1)
\frac{\Lambda^{2\varepsilon}(x)x_1^2}{(m_1(x_1))^2}\\&\times\sum_{y_2\in\Vc_2}m_2(y_2)
\frac{\sin^2(\frac{\gamma(x,(x_1,y_2))}{2})}{1+\mu(x_1,y_2)}\Big(1+\deg_{\Gc}(x_1+1,y_2)\Big)^{-1}
\\&
\times\Big|\Big(1+\deg_{\Gc}(x_1+1,y_2)\Big)^{\frac{1}{2}}f(x_1+1,y_2)\Big|^2
\frac{1}{m_2(y_2)}\\&\times\sum_{x_2\in\Vc_2}\frac{\Ec_2^2(x_2,y_2)m_2(x_2)}{(m_2(x_2))^2(1+\mu(x_1,y_2))}\Big(1+\deg_{\Gc}(x_1,x_2)\Big)^{-1}
\\
&\leq8c\sum_{x_1\in\Vc_1}m_1(x_1)\frac{(\Lambda|_{\Gr^*})^{2\varepsilon}(x)x_1^2}{(m_1(x_1))^2}\sum_{y_2\in\Vc_2}m_2(y_2)
\frac{\sin^2(\frac{\gamma(x,(x_1,y_2))}{2})}{1+\mu(x_1,y_2)}\\&\times\Big|\Big(1+\deg_{\Gc}(x_1+1,y_2)\Big)^{\frac{1}{2}}f(x_1+1,y_2)\Big|^2
\\
&\times\frac{1}{m_2(y_2)}\sum_{x_2\in\Vc_2}\frac{\Ec_2^2(x_2,y_2)}{m_2(x_2)}\Big(1+\deg_{\Gc}(x_1,x_2)\Big)^{-1}\Big(1+\deg_{\Gc}(x_1+1,y_2)\Big)^{-1}
\\
&\leq8c\sum_{x_1\in\Vc_1}m_1(x_1)\frac{(\Lambda|_{\Gr^*})^{2\varepsilon}(x)x_1^2}{(m_1(x_1))^2}\sum_{y_2\in\Vc_2}m_2(y_2)
\frac{\sin^2(\frac{\gamma(x,(x_1,y_2))}{2})}{1+\mu(x_1,y_2)}\\&\times\Big|\Big(1+\deg_{\Gc}(x_1+1,y_2)\Big)^{\frac{1}{2}}f(x_1+1,y_2)\Big|^2
\frac{1}{m_2(y_2)}
\sum_{x_2\in\Vc_2}\frac{\Ec_2^2(x_2,y_2)}{m_2(x_2)}\\
&\quad\times\min\Big(1,\frac{m_2(y_2)}{\deg_{\Gc}(x_1+1,x_2)}\Big)\min\Big(1,\frac{m_2(x_2)}{\deg_{\Gc}(x_1,x_2)}\Big)
\\
&\leq 8c(e+e^{-1})^{-2}\sum_{x_1\in\Vc_1}m_1(x_1)\frac{(\Lambda|_{\Gr^*})^{2\varepsilon}(x)x_1^2}{(m_1(x_1))^2}\sum_{y_2\in\Vc_2}m_2(y_2)\deg_{\Gc_2}
\\&\times\frac{\sin^2(\frac{\gamma(x,(x_1,y_2))}{2})}{1+\mu(x_1,y_2)}(y_2)\Big|\Big(1+\deg_{\Gc}(x_1+1,y_2)\Big)^{\frac{1}{2}}f(x_1+1,y_2)\Big|^2
\\&\times\frac{\min\Big(e+e^{-1},m_2(y_2)\Big)}{m_2(y_2)}
\sum_{x_2\in\Vc_2}\frac{\Ec_2(x_2,y_2)}{m_2(x_2)}\min\Big(e+e^{-1},m_2(x_2)\Big).
\end{align*}
o ensure the boundedness of \begin{align}\label{B}\Big\|(\Lambda|_{\Gr^*})^{\varepsilon}(\cdot)Bf\Big\|_{\Gr^*}^2=\Big\|\Big(1+\deg_{\Gc}(\cdot)\Big)^{-\frac{1}{2}}\Lambda|_{\Gr^*}^{\varepsilon}(\cdot)Bf\Big\|^2, \end{align}
it is enough to assume the the following hypothesis:
\begin{align}\label{H8'}&\sup_{(x_1,x_2)\in\Vc}\sup_{y_2\in\Vc_2}\Bigg|\sin(\frac{\gamma(x,(x_1,y_2))}{2})\Bigg|
\deg_{\Gc_2}(y_2)m_2^{\frac{1}{2}}(y_2)\\&\nonumber\times\min\Bigg((e+e^{-1}
\sum_{x_2\in\Vc_2}\frac{m_2^{\frac{1}{2}}(y_2)}{m_2(x_2)},1,\Big(\sum_{x_2\in\Vc_2}\frac{m_2^2(y_2)}{m_2(x_2)}\Big)
^{\frac{1}{2}},m_2(y_2)\Bigg),
\end{align} is also bounded. Specifically, if $m_2(\cdot)$ is bounded, then it follows that \eqref{B} will remain bounded as well. Otherwise, if $m_2(y_2)\rightarrow\infty, \hbox{ when } |y_2|\rightarrow\infty $, we can reformulate the hypothesis \eqref{H8'} in a more comprehensible manner the Hypothesis $(H_5)$.
At this juncture, we shall now focus our attention on the two first term of
$$\Big\|(\Lambda|_{\Gr^*})^{\varepsilon}[D_2,Q_1U^*]f\Big\|_{\Gr^*}=\Big\|\Big(1+\deg_{\Gc}(\cdot)\Big)^{-\frac{1}{2}}(\Lambda|_{\Gr^*})[D_2,Q_1U^*]f\Big\|.$$
Since the assertions $(H_1)$, $(H_2)$, $(H_5)$ and $(H_6)$ hold true then there exists an integer $c$, such that:
\begin{align*}&\sum_{x\in\Vc}m(x)\Big(1+\deg_{\Gc}(x)\Big)^{-1}\Bigg|(\Lambda|_{\Gr^*})^{\varepsilon}(x)\Bigg(\frac{1}{m(x)}x_1\Ec_1(x_1,x_1+1)
\frac{1-e^{\rmi \gamma(x,x+1)}}{\sqrt{(1+\mu(x))(1+\mu(x+1))}}
\\
&-\frac{1}{m(x+1)}x_1\Ec_1(x_1+1,x_1+2)
\frac{1-e^{\rmi \gamma(x+1,x+2)}}{\sqrt{(1+\mu(x+1))(1+\mu(x+2))}}
\Bigg)f(x+2)\Bigg|^2
\\
&=\sum_{x\in\Vc}m(x)\Big(1+\deg_{\Gc}(x)\Big)^{-1}\Bigg|(\Lambda|_{\Gr^*})^{\varepsilon}(x)x_1\otimes\un_{\Vc_2}\frac{e^{\frac{1}{2}}}{\sqrt{1+\mu(x+1)}}\Bigg|^2
\\&\times\Bigg|\Bigg(
\frac{1-e^{\rmi \gamma(x,x+1)}}{\sqrt{1+\mu(x)}}-\frac{1-e^{\rmi \gamma(x+1,x+2)}}{\sqrt{1+\mu(x+2)}}\Bigg)\Bigg|^2\\&\times\Bigg|\Big(1+\deg_{\Gc}(x+2)\Big)^{-\frac{1}{2}}
\Big(1+\deg_{\Gc}(x+2)\Big)^{\frac{1}{2}}f(x+2)\Bigg|^2
\\
&=\sum_{x\in\Vc}m(x)\Big(1+\deg_{\Gc}(x)\Big)^{-1}\Big(1+\deg_{\Gc}(x+2)\Big)^{-1}
\Bigg|(\Lambda|_{\Gr^*})^{\varepsilon}(x)x_1\otimes\un_{\Vc_2}\frac{e^\frac{1}{2}}{\sqrt{1+\mu(x+1)}}\Bigg|^{2}
&\hspace*{-1.37cm}\\&\times\Bigg|\Bigg(\Big(\frac{e^{\rmi \gamma(x+1,x+2)}}{\sqrt{1+\mu(x+2)}}
-\frac{e^{\rmi \gamma(x,x+1)}}{\sqrt{1+\mu(x)}}\Big)+\Big(\frac{1}{\sqrt{1+\mu(x)}}
-\frac{1}{\sqrt{1+\mu(x+2)}}\Big)\Bigg)\Bigg|^{2}\\&\times\Bigg|\Big(1+\deg_{\Gc}(x+2)\Big)^{\frac{1}{2}}f(x+2)\Bigg|^2
\\&
\leq C\sum_{x\in\Vc}m(x)\Big(1+\deg_{\Gc}(x)\Big)^{-2}
\Bigg|(\Lambda|_{\Gr^*})^{\varepsilon}(x)x_1\otimes\un_{\Vc_2}\frac{e^\frac{1}{2}}{\sqrt{1+\mu(x+1)}}\Bigg|^{2}
&\hspace*{-1.37cm}\\&\times\Bigg|\Bigg(\Big(\frac{e^{\rmi \gamma(x+1,x+2)}}{\sqrt{1+\mu(x+2)}}
-\frac{e^{\rmi \gamma(x,x+1)}}{\sqrt{1+\mu(x)}}\Big)+\Big(\frac{1}{\sqrt{1+\mu(x)}}
-\frac{1}{\sqrt{1+\mu(x+2)}}\Big)\Bigg)\Bigg|^{2}\\&\times\Bigg|\Big(1+\deg_{\Gc}(x+2)\Big)^{\frac{1}{2}}f(x+2)\Bigg|^2
\\
&\leq2C\sum_{x\in\Vc}m(x)\Big(1+\deg_{\Gc}(x)\Big)^{-2}\Bigg|(\Lambda|_{\Gr^*})^{\varepsilon}(x)x_1\otimes\un_{\Vc_2}\frac{1}{\sqrt{1+\mu(x+1)}}
\\&\Bigg(\frac{e^{\rmi \gamma(x+1,x+2)}\sqrt{1+\mu(x)}-e^{\rmi \gamma(x,x+1)}\sqrt{1+\mu(x+2)}}{\sqrt{(1+\mu(x))(1+\mu(x+2)}}\Bigg)\Bigg|^2\Bigg|\Big(1+\deg_{\Gc}(x+2)\Big)^{\frac{1}{2}}f(x+2)\Bigg|^2
\\
&+2C\sum_{x\in\Vc}m(x)\Bigg|(\Lambda|_{\Gr^*})^{\varepsilon}(x)x_1\otimes\un_{\Vc_2}
\frac{\sqrt{1+\mu(x+2)}-\sqrt{1+\mu(x)}}{\sqrt{(1+\mu(x+1))(1+\mu(x))(1+\mu(x+2))}}\Bigg|^2
\\&\times\Bigg|\Big(1+\deg_{\Gc}(x+2)\Big)^{\frac{1}{2}}f(x+2)\Bigg|^2
\\
&\leq2C\sum_{x\in\Vc}m(x)\Big(1+\deg_{\Gc}(x)\Big)^{-2}\Bigg|\Big(1+\deg_{\Gc}(x+2)\Big)^{\frac{1}{2}}f(x+2)\Bigg|^2
\\&\hspace*{-1.4cm}\times\Bigg|(\Lambda|_{\Gr^*})^{\varepsilon}(x)x_1\otimes\un_{\Vc_2}\frac{1}{\sqrt{1+\mu(x+1)}}
\Bigg(\frac{e^{\rmi \gamma(x+1,x+2)}\sqrt{1+\mu(x)}-e^{\rmi \gamma(x,x+1)}\sqrt{1+\mu(x+2)}}{\sqrt{(1+\mu(x))(1+\mu(x+2))}}\Bigg)\Bigg|^2
\\
&+2C\sum_{x\in\Vc}m(x)\Big(1+\deg_{\Gc}(x)\Big)^{-2}\Bigg|\Big(1+\deg_{\Gc}(x+2)\Big)^{\frac{1}{2}}f(x+2)\Bigg|^2
\\&\times\Bigg|(\Lambda|_{\Gr^*})^{\varepsilon}(x)x_1\otimes\un_{\Vc_2}
\frac{\sqrt{1+\mu(x+2)}-\sqrt{1+\mu(x)}}{\sqrt{(1+\mu(x+1))(1+\mu(x))(1+\mu(x+2))}}\Bigg|^2
\\
&\leq4C\sum_{x\in\Vc}m(x)\Big(1+\deg_{\Gc}(x)\Big)^{-2}\Bigg|\Big(1+\deg_{\Gc}(x+2)\Big)^{\frac{1}{2}}f(x+2)\Bigg|^2
\\&\times\Bigg|(\Lambda|_{\Gr^*})^{\varepsilon}(x)x_1\otimes\un_{\Vc_2}
\frac{e^{\rmi \gamma(x+1,x+2)}-e^{\rmi \gamma(x,x+1)}}{\sqrt{(1+\mu(x+1))(1+\mu(x))}}\Bigg|^2
\\
&+4C\sum_{x\in\Vc}m(x)\Big(1+\deg_{\Gc}(x)\Big)^{-2}\Bigg|\Big(1+\deg_{\Gc}(x+2)\Big)^{\frac{1}{2}}f(x+2)\Bigg|^2
\\&\times\Bigg|(\Lambda|_{\Gr^*})^{\varepsilon}(x)x_1\otimes\un_{\Vc_2}
\frac{e^{\rmi \gamma(x+1,x+2)}\Big(\sqrt{1+\mu(x)}-\sqrt{1+\mu(x+2)}\Big)}
{\sqrt{(1+\mu(x))(1+\mu(x+2))(1+\mu(x+1))}}\Bigg|^2
\\
&+2C\sum_{x\in\Vc}m(x)\Big(1+\deg_{\Gc}(x)\Big)^{-2}\Bigg|\Big(1+\deg_{\Gc}(x+2)\Big)^{\frac{1}{2}}f(x+2)\Bigg|^2
\\&\times\Bigg|(\Lambda|_{\Gr^*})^{\varepsilon}(x)x_1\otimes\un_{\Vc_2}
\frac{\sqrt{1+\mu(x+2)}-\sqrt{1+\mu(x)}}
{\sqrt{(1+\mu(x))(1+\mu(x+2))(1+\mu(x+1))}}\Bigg|^2
\\
&\leq4C\sum_{x\in\Vc}m(x)\Bigg|\Big(1+\deg_{\Gc}(x+2)\Big)^{\frac{1}{2}}f(x+2)\Bigg|^2
\\&\times\Bigg|(\Lambda|_{\Gr^*})^{\varepsilon}(x)x_1\otimes\un_{\Vc_2}
\frac{\Big(1+\deg_{\Gc}(x)\Big)^{-1}\Big(e^{\rmi \gamma(x+1,x+2)}-e^{\rmi \gamma(x,x+1)}\Big)}{\sqrt{(1+\mu(x+1))(1+\mu(x))}}\Bigg|^2
\\
&+6C\sum_{x\in\Vc}m(x)\Bigg|\Big(1+\deg_{\Gc}(x+2)\Big)^{\frac{1}{2}}f(x+2)\Bigg|^2\Bigg|\frac{\Big(1+\deg_{\Gc}(x)\Big)^{-1}\Big(\mu(x)-\mu(x+2)\Big)}
{\sqrt{1+\mu(x)}+\sqrt{1+\mu(x+2)}}\Bigg|^2
\\&\times\Bigg|(\Lambda|_{\Gr^*})^{\varepsilon}(x)x_1\otimes\un_{\Vc_2}
\frac{1}{\sqrt{(1+\mu(x))(1+\mu(x+2))(1+\mu(x+1))}}\Bigg|^2
\\&
\leq c\|f\|_{\mathcal{G}}^2.
\end{align*}
For sufficiently large $|(x,y)|$, it holds that $|\gamma(x,y)|<\frac{\pi}{4}$.
Also, it is established that for $|t|\in[0,\frac{\pi}{2}]$, $\frac{2}{\pi}|t|\leq\sin|t|\leq |t|$.
Consequently, the hypothesis
\begin{align*}\sup_{(x_1,x_2)\in\Vc}\langle x_1\rangle^{1+\varepsilon} \bigg|\sin\Big(\frac{\gamma((x_1,x_2),(x_1+1,x_2))-\gamma((x_1,x_2),(x_1-1,x_2)}{2}\Big)\bigg|<\infty\end{align*}
 is equivalent to hypothesis $(H_1)$.
In the same way, we deal with \eqref{e:1} and by using the hypothesis $(H_7)$, we demonstrate the boundedness of $\|\Big(1+\deg_{\Gc}(\cdot)\Big)^{-\frac{1}{2}}\Lambda^{\varepsilon}[D_1,Q_1U^*]\|$.
Similarly, we treat the other terms. By density, there exists $c_{\varepsilon}>0$ such that: $$\|(\Lambda|_{\Gr^{*}})^{\varepsilon}[\widetilde\Delta_{\Gc_{\xi, \mu, \gamma}^{}}- \Delta_\Gc, \Ac_{\Gc}]f\|^2_{\Gr^{*}}\leq c\|f\|^2_{\Gr}.$$

Finally,  by applying  \cite[Proposition 7.5.7]{ABG} where the hypotheses are verified in Proposition \ref{p:hypothesis}, we find that  $[\widetilde\Delta_{\Gc_{\xi, \mu, \gamma}^{}}- \Delta_\Gc,\rmi\Ac_{\Gc}]_{\circ}$ is in $\mathcal{C}^{1}(\Ac,\Gr,\Gr^{*})$ by taking $\varepsilon=0$ and  in $\mathcal{C}^{1,1}(\Ac,\Gr,\Gr^{*})$ considering $\varepsilon\in ]0,\epsilon]$. \qed

\subsection{Proof of the main result}
We finish with the proof of Theorem \ref{t:LAP}.
Note that the conjugate operator has $\un_{\Vc_2}$ in its definition. Therefore, we have without any changes in the proof
consider a more general graph (see eg. \cite{GT}) given by $\sum_{z\in\mathfrak{J}}\delta_{y,z}$ with $\mathfrak{I}\in V_2$ and non empty.

First $\Delta_{\Gc_{\xi,\mu,\theta}}+V(\cdot)\in\Cc^{1,1}(\Ac_{\Gc_{\xi,\mu,\theta}};\Gr,\Gr^*)$ because $\Delta_{\Gc_{\xi,\mu,\theta}}\in\Cc^{1,1}(\Ac_{\Gc_{\xi,\mu,\theta}};\Gr,\Gr^*)$
and $V(\cdot)\in\Cc^{1,1}(\Ac_{\Gc_{\xi,\mu,\theta}};\Gr,\Gr^*)$ by Proposition \ref{C1,1fun} and Lemma \ref{l:Vfun}.
In particular, we have that $\Delta_{\Gc_{\xi,\mu,\theta}}\in\Cc^1_u(\Ac_{\Gc_{\xi,\mu,\theta}};\Gr,\Gr^*)$, see \cite{ABG}.

Then, using Proposition \ref{p:Mourrem2}, Proposition \ref{p:com}, and by \cite[Theorem 7.2.9]{ABG}, 
there exist $c>0$, a compact operator $K$ such that \begin{align*}E_{\Ic}(\Delta_{\Gc_{\xi,\mu,\theta}}+V(\cdot))[\Delta_{\Gc_{\xi,\mu,\theta}}&+V(\cdot), \rmi \Ac_{\Gc_{\xi,\mu,\theta}}]_\circ E_{\Ic}(\Delta_{\Gc_{\xi,\mu,\theta}}+V(\cdot))
\\
&\geq cE_{\Ic}(\Delta_{\Gc_{\xi,\mu,\theta}}+V(\cdot))+K. \end{align*}
Now, we turn to points $(3)$. It is enough to obtain them with $s\in(1/2,1)$. We apply \cite[Proposition 7.5.6]{ABG}, we obtain:
$$\lim_{\rho\rightarrow 0^{+}}\sup_{\lambda\in[a,b]}\|\langle\Ac_{\overline{\Gc},\Gr}\rangle^{-s}(H-\lambda-\rmi\rho)^{-1}\langle \Ac_{\overline{\Gc},\Gr^{*}}\rangle^{-s}\|_{\Bc(\ell^2(\Vc, m_\mu))} \mbox{ is finite.}$$
In addition, in the norm topology of bounded operators, the boundary values of the
resolvent:
\[ [a,b] \ni\lambda\mapsto\lim_{\rho\to0^{\pm}}\langle \Ac_{\overline{\Gc},\Gr}\rangle^{-s}(H-\lambda-\rmi\rho)^{-1}\langle \Ac_{\overline{\Gc},\Gr^{*}}\rangle^{-s} \mbox{ exists and is continuous},\]
where $[a,b]$ is included in $\mathbb{R}\setminus\left(\kappa\cup\sigma_{\rm p}(H)\right)$.\\
Using Proposition \ref{p:hypothesis} and Lemma \ref{l:interpo}, we obtain
\begin{align*}\lim_{\rho\rightarrow0^{\pm}}\sup_{\lambda\in[a,b]}\|(\Lambda|_{\Gr})^{-s}
(\Delta_{\Gc_{\xi,\mu,\theta}}+V(\cdot)-\lambda-\rmi\rho)^{-1}
(\Lambda|_{\Gr^*})^{-s}\|_{\Bc(\ell^2(\Vc, m_\mu)},\end{align*}
exists and finite.  Finally The Point $(4)$  follows from Point  $(3)$ by standard ways, e.g.
\cite[Theorem XII.25, vol.4]{RS}. \qed

\end{document}